\documentclass[12pt, letterpaper, oneside]{article}
\usepackage{todonotes}
\usepackage[utf8]{inputenc}
\usepackage{amsmath,bm}
\usepackage{graphicx}
\usepackage[legalpaper, margin=1in]{geometry}
\usepackage{mathtools}
\usepackage{breqn}
\usepackage{relsize}
\usepackage{siunitx}
\usepackage{amssymb}
\usepackage{gensymb}
\usepackage{authblk}
\usepackage[backend=bibtex,style=phys]{biblatex}

\title{Shear rheology of a dilute emulsion of ferrofluid droplets dispersed in a non-magnetizable carrier fluid under the influence of a uniform magnetic field}

\author[1]{P. Capobianchi\footnote{Corresponding author. Email: paolo.capobianchi@strath.ac.uk}}
\author[1]{M. Lappa}
\author[1]{M.S.N. Oliveira}
\author[2]{F.T. Pinho}
\affil[1]{James Weir Fluid Lab, Department of Mechanical and Aerospace Engineering, University of Strathclyde, 75 Montrose Street, Glasgow G1 1XJ, UK}
\affil[2]{CEFT, Departamento de Engenharia Mecânica, Faculdade de Engenharia da Universidade do Porto, Rua Dr. Roberto Frias, 4200-465 Porto, Portugal}
\date{}

\begin{document}

\maketitle

\begin{abstract}
\noindent The effect of a spatially uniform magnetic field on the shear rheology of a dilute emulsion of monodispersed ferrofluid droplets, immersed in a non-magnetizable immiscible fluid, is investigated using direct numerical simulations. The direction of the applied magnetic field is normal to the shear flow direction. The droplets extra stress tensor arising from the presence of interfacial forces of magnetic nature is modeled on the basis of the seminal work of G. K. Batchelor, J. Fluid Mech., 41.3 (1970) under the assumptions of a linearly magnetizable ferrofluid phase and negligible inertia. The results show that even relatively small magnetic fields can have significant consequences on the rheological properties of the emulsion due to the magnetic forces that contribute to deform and orient the droplets towards the direction of the applied magnetic vector. In particular, {we have} observed an increase of the effective (bulk) viscosity and a reversal of the sign of the two normal stress differences with respect to the case without magnetic field {for those conditions where the magnetic force prevails over the shearing force}. Comparisons between the results of our model with a direct integration of the viscous stress have provided an indication of its reliability to predict the effective viscosity of the suspension. Moreover, this latter quantity {has been} found to {behave as} a monotonic increasing function of the applied magnetic field for constant shearing flows (``magneto-thickening'' behaviour), which allowed us to {infer} a simple constitutive equation describing the emulsion viscosity.
\end{abstract}

\section{\label{introduction}Introduction}

Heterogeneous mixtures of small particles of various types such as solid particles, bubbles and droplets, dispersed in a carrier fluid are widespread in many industrial, chemical and biological processes. Typical applications can be encountered, for instance, in oil and gas industry, mining processes, in electronic devices, in biomedical applications and food industry. Owing to their great scientific and industrial relevance, suspensions have been the object of extensive studies over the past decades. Yet, due to their variety and the complexity of their rheology under a wide range of conditions (e.g., appearance of inter-particle interactions, presence of additional constraints like electric and magnetic fields, or different characteristics of the dispersed phase), suspensions are still actively investigated today. 

Provided the length scale of the applied flow is large compared with the mean particle dimension, suspensions may be regarded as homogeneous fluids in some instances and their rheological properties can be evaluated using standard rheometric flows, i.e., steady shear, extensional and small amplitude oscillatory shear flow. In a steady shear-flow experiment, the response of the system is completely characterised by three independent parameters: the shear viscosity, $\eta=\Sigma_{xy}/\dot{\gamma}$, and the two normal stress differences, $N_1=\Sigma_{xx}-\Sigma_{yy}$ and $N_2=\Sigma_{yy}-\Sigma_{zz}$ (or their coefficients, $\Psi_1=N_1/\dot{\gamma}^2$, $\Psi_2=N_2/\dot{\gamma}^2$), where $\boldsymbol{\Sigma}$ is the total stress tensor ($\Sigma_{xy}$ is the shear component, while $\Sigma_{xx}$, $\Sigma_{yy}$ and $\Sigma_{zz}$ are the three normal components) and $\dot{\gamma}$ is the rate of deformation.

Studies on the rheology of suspensions can be traced back to the seminal work of A. Einstein \cite{einstein1906,einstein1911}. Einstein showed that the effective viscosity of a dilute suspension of rigid Brownian spheres can be described as $\eta_{e}=\eta\left(1+2.5\phi \right)$, where $\eta$ is the viscosity of the carrier fluid and $\phi$ is the volume fraction of the dispersed phase. Later, G.I. Taylor \cite{taylor1932} obtained an analogous expression for the effective viscosity of a dilute emulsion derived in the framework of small deformation theory, $\eta_{e}=\eta\left[1+2.5\phi(\lambda+2/5)/(\lambda+1) \right]$, where $\lambda$ is the drop-to-continuous phase viscosity ratio. In the limiting case $\lambda \to \infty$, the emulsion behaves like a dilute suspension of rigid spheres dispersed in a viscous fluid, and Taylor's equation reduces to Einstein's equation. In the opposite case, i.e., for $\lambda \to 0$, the emulsion can be regarded as a foam-like material and the expression for its effective viscosity becomes $\eta_{e}=\eta\left( 1+\phi \right)$ (cf., e.g. Derkach \cite{derkach2009}). 

The abovementioned theories predict a constant (Newtonian) shear viscosity, however it is well-known that suspensions can exhibit different non-Newtonian behaviour. In the case of hard sphere colloidal suspensions (HS, in the following), the rheological properties are essentially determined by the volume fraction of the dispersed phase, $\phi$, and by the Péclet number, $Pe=\tau_B \dot{\gamma}$, where $\tau_B$ is the Brownian time scale, i.e., the time required for a free particle to diffuse its own radius \cite{guyetal2015}. In the Brownian regime, $Pe\ll1$ and these suspensions exhibit a Newtonian-like behaviour consistent with Einstein's equation. Shear-thinning effects start to become appreciable at $Pe\approx1$, followed by a second Newtonian regime at $Pe\gg1$ with a viscosity that finally diverges (shear-thickening regime) at random close packing, $\phi_{RCP}\approx0.64$ (cf. Ref.\cite{mewis_wagner_2011}). Moreover, normal stress differences can also be detected in simple shearing experiments with hard sphere suspensions \cite{laun1984,zarraga2000,pan2017}. While these suspensions are usually characterised by a negative first normal stress difference for moderately dense regimes, transition from negative to positive $N_1$ can be observed at high shear rates for very dense regimes \cite{marietal2014,boromandetal2018}.

A particular type of dispersion of hard particles which finds a multitude of practical and scientific implications are ferrofluids (FFs). These fluids are colloidal suspensions of nanosized (typically $d_p\lesssim10\,\si{\nano\metre}$, where $d_p$ is the particle diameter) superparamagnetic particles dispersed in a continuous fluid. Without the presence of a magnetic field, the particles remain randomly dispersed in the carrier phase due to Brownian effects and FFs can be regarded as regular nanofluids. For dilute suspensions, FFs essentially exhibit a Newtonian behaviour, while for sufficiently large concentrations, shear-thinning effects may become evident (see, e.g., Ref.\cite{shahrivar2019}). In the presence of magnetic fields, however, a variety of different non-Newtonian behaviour may appear. {{In this regard, we can distinguish between the ideal scenario in which inter-particle interactions are considered absent (``non-interacting'' (NI) ferrofluid models) and the case in which particle-particle interactions are non-negligible and chain-like aggregates may appear (see, e.g., \cite{Ilg2009} for a detailed overview of the subject). In the absence of particle interactions (ideal ferrofluid, in the following)}, the rheological behaviour of the material is essentially dictated by the response of the particles to the magnetic field, in addition to Brownian and hydrodynamics effects. In such conditions, the presence of a magnetic moment imposes a constraint on the rotation of each particle (that would otherwise be free to rotate under the effect of the vorticity component of the flow). As a result, an additional viscous dissipation appears, which ultimately leads to an increase in the suspension viscosity (magnetoviscous effect \cite{ROSENSWEIG1969680,mctague1969}), usually accounted for with an additional ``rotational viscosity'', $\eta_r$. Several theories have succeeded in describing this effect for non-interacting particles. Worth mentioning is the early macroscopic (phenomenological) theory of Shilomis \cite{shliomis1972} and the subsequent microscopic theories of Brenner and Weissman \cite{BRENNER1972499} and Martsenyuk et al. \cite{martsenyuk1974}. Specifically, Martsenyuk et al. \cite{martsenyuk1974} derived an expression for the rotational viscosity which is proportional to the particle volume fraction, $\phi$, and the Langevin parameter, i.e., $\eta_r\left(\beta \right)=\frac{3}{2}\eta\phi{\beta L\left(\beta \right)}/\left(\beta-L\left(\beta \right)\right)$, where $\beta=mH/k_BT$ is the Langevin parameter in which $m$ is the magnetization moment, $H$ is the intensity of the magnetic field, $k_B$ is the Boltzmann constant and $T$ is the absolute temperature, whereas $L\left(\beta \right)$ represents the Langevin function. Hence, it can be observed that in the absence of magnetic field, $\eta_r\left(0 \right)=0$, while, on the contrary, if the field is strong enough to prevent completely particle rotation, $\eta_r\left(\infty \right)=\frac{3}{2}\eta\phi$. Since the volume fraction of spherical particles is $\phi\approx0.74$ near the densest close packing, the maximum rotational viscosity predictable by the theory of Martsenyuk et al. \cite{martsenyuk1974} is $\eta_{r,max}\approx1.1\eta$.} 

{Despite the success of these theories on capturing the magnetoviscous effect in very dilute ferrofluids (e.g., see the comparison between experiments and the theory of Martsenyuk et al. \cite{martsenyuk1974} reported in McTague \cite{mctague1969}), the agreement with experiments for moderately to highly concentrated suspension is unsatisfactory (a relative increment in viscosity of about 200\% with respect to the continuous liquid phase viscosity $\eta$ was already detected in the early observation of Rosensweig et al. \cite{ROSENSWEIG1969680}). Such discrepancies might be justified considering the occurrence of interactions between particles. Indeed, it is well-known that upon the application of a magnetic field, dipolar and steric interactions may promote the formation of chain-like aggregates which, on the one hand contribute to enhance the aforementioned magnetoviscous effect, and, on the other hand, confer additional rheological attributes that are typically encountered in non-Newtonian fluids, such as shear-thinning effects, a yield stress \cite{Zubarev_2006} and viscoelastic effects, namely normal stress differences in simple shearing flow \cite{Zubarev_1992,Odenbach1999}}.

Similarly to hard sphere suspensions, emulsions also exhibit several non-Newtonian features. Contrarily to HS, however, the deformability of the dispersed phase introduces additional complexity {into} the system, originating rheological properties that are intimately connected to the morphological microstructure of the droplets evolving under the effect of a flow. Deformation-induced shear-thinning is a distinguishing mark of these systems (see, for instance Ref. \cite{derkach2009}). Moreover, unlike hard sphere suspensions, emulsions are usually characterised by a positive first normal stress difference \cite{loewenberghinch1996,zinchenko2003LargescaleSO}, a signature of {viscoelasticity}, although some authors have reported a reversal in the sign of $N_1$ attributed to the presence of inertial effects \cite{li_sarkar2005}.

After the early efforts of Taylor \cite{taylor1932,taylor1934}, many authors attempted to unveil the richness of the physics involved in the dynamics of emulsions evolving under different flow conditions. The amount of literature regarding this subject is indeed very vast. Oldroyd \cite{oldroyd1953} derived a linear viscoelastic constitutive equation for time-dependent flows, corroborated by expressions for the relaxation and retardation times of the fluid proportional to the droplet capillary time scale. Later, Schowalter et al. \cite{SCHOWALTER1968152} investigated the behaviour of a drop under steady shear adopting a first-order perturbation method and determined a positive $N_1$ and a negative $N_2$, both proportional to the square of the rate of deformation, $\dot{\gamma}$. Frankel and Acrivos \cite{frankelacrivos1970} generalized the theory of Schowalter et al. \cite{SCHOWALTER1968152} for a time-dependent shearing flow for a dilute emulsion and obtained the expression for the stress tensor. Subsequently, Cox \cite{cox1969} provided a solution for the drop shape in a rather general time-dependent creeping flow.

In addition to these works, which were specifically aimed at determining the flow field and the morphological configuration of a single drop under certain flow conditions, other authors developed theories aimed at describing the rheological properties of suspensions in terms of average particle interfacial stress \cite{batchelor_1970,MELLEMA1983286,onuki1987}. In particular, Batchelor \cite{batchelor_1970} obtained an expression for the bulk stress of a suspension of particles of generic shape and constitution (solid particles, drops, capsules, etc.) in Newtonian fluids in the absence of external body forces, while allowing for the presence of couples exerted on those particles. Apart from these limiting assumptions, the derivation of Batchelor \cite{batchelor_1970} is rather general and can, in principle, be adopted for any type of suspension regardless the concentration of the dispersed phase. Later, Choi and Schowalter \cite{choi1975} determined constitutive equations for non-dilute suspensions adopting the definition of interfacial stress tensor given in Batchelor \cite{batchelor_1970}. 

More recently, various authors have approached the problem from a phenomenological perspective and succeeded in obtaining accurate predictions for both droplet conformation and rheological behaviour of dilute emulsions \cite{MAFFETTONE1998227,garmelaetal2001,yu_bousmina2002}.

Aside from viscous (and possibly inertial \cite{li_sarkar2005}) effects arising from the presence of an imposed flow, the configuration of the dispersed phase can also be altered by additional stresses of different nature, such as electric fields  (e.g., see \cite{taylor1966,torza1971,vlahovska_2011}) and magnetic fields \cite{CAPOBIANCHI2018313,hassan2018,cunha2018,cunha2020,ishida2020}. In the latter case, at least one phase must be composed of a magnetizable material. Cunha et al. \cite{cunha2020} derived a model for the interfacial stress tensor developing in the presence of uniform magnetic fields. Then, they applied their model to the two-dimensional problem of a dilute emulsion composed of ferrofluid droplets surrounded by a non-magnetizable fluid under a steady shearing flow and a uniform magnetic field acting both in the normal and parallel directions with respect to the imposed flow. They calculated the effective viscosity by integrating the viscous stresses at the wall and found good agreement with the prediction of their model. Moreover, in both flow conditions they found positive first normal stress differences. {More recently, Ishida and Matsunaga \cite{ishida2020} have also proposed a model for the rheology of a dilute emulsion of ferrofluid droplets dispersed in a non-magnetizable medium approaching the problem considering two- and three-dimensional configurations and uniform magnetic fields applied along each of the three coordinate directions. They observed a reversal of normal stress differences with respect to non-magnetic configuration when the magnetic field was parallel to the direction of the vorticity vector. Regarding the shear viscosity, their two-dimensional calculations have shown a general good agreement with the finding of Cunha et al. \cite{cunha2020}}.

To the best of our knowledge, the abovementioned {works} of Cunha et al. \cite{cunha2018,cunha2020} {and Ishida and Matsunaga \cite{ishida2020} are} the only aimed at investigating the rheology of emulsions in the presence of a ferrofluid phase and an imposed magnetic field. {Previous works, on the other hand, have been devoted to the study of non-rheological properties such as emulsion magnetic permeability, for emulsions of ferrofluid drops in non-magnetizable fluids (e.g., see Refs. \cite{ivanovetal2011,ivanovetal2013,Zakinyan2011}) as well as for inverse emulsions, i.e., for non-magnetizable drops surrounded by a ferrofluid (cf. Refs. \cite{Subbotin2018MagneticPO,SUBBOTIN2020166524}), the formation of chained structures of ferrofluid droplets, \cite{Jain:ay5476} and the effect of these structures on the emulsion electrical properties \cite{Zakinyan2014}. Finally, it is worth mentioning the recent work of Zakinyan and Zakinyan \cite{ZAKINYAN2020112347} who succeed on producing an emulsion of ferrofluid microdrops using a rotating magnetic field and showed that the resulting magnetic torque of the emulsion can be enhanced with respect that observable in the pure ferrofluid.} 

{From this brief account, it clearly emerges that ferrofluid emulsion show the potential for being employed in a wide range of novel scientific and engineering applications owing to the possibility to ``tune'' their mechanical and electromagnetic properties ad hoc with the application of opportune magnetic fields. Nevertheless, if some of the aspects related to the electromagnetic properties of these systems have been already studied theoretically end experimentally in certain detail, contrarily, works specifically aimed at investigating their rheological properties are relatively scarce. In the present work, therefore,} a model based on the theory of Batchelor \cite{batchelor_1970} is developed anew {following a different route from those adopted in the aforementioned works of Cunha et al. \cite{cunha2020} and Ishida and Matsunaga \cite{ishida2020}. Qualitative comparisons with the previous findings reported in \cite{cunha2020,ishida2020} provided evidence of agreement between different models on predicting shear stresses. On the contrary, discrepancies in terms of normal stress differences might be expected since the present approach predicted the reversal of the normal stress differences, while, for flow conditions comparable to ours, in the above mentioned works of Cunha et al. \cite{cunha2020} and Ishida and Matsunaga \cite{ishida2020} this occurrence was not observed. Finally, a quantitative comparison with our results and those obtained with the adoption of the model developed by Cunha et al. \cite{cunha2020} is is also provided.}

\section{\label{problem_statement}Problem formulation}

\begin{figure}[h!]
\centering
\includegraphics[width=0.7\textwidth]{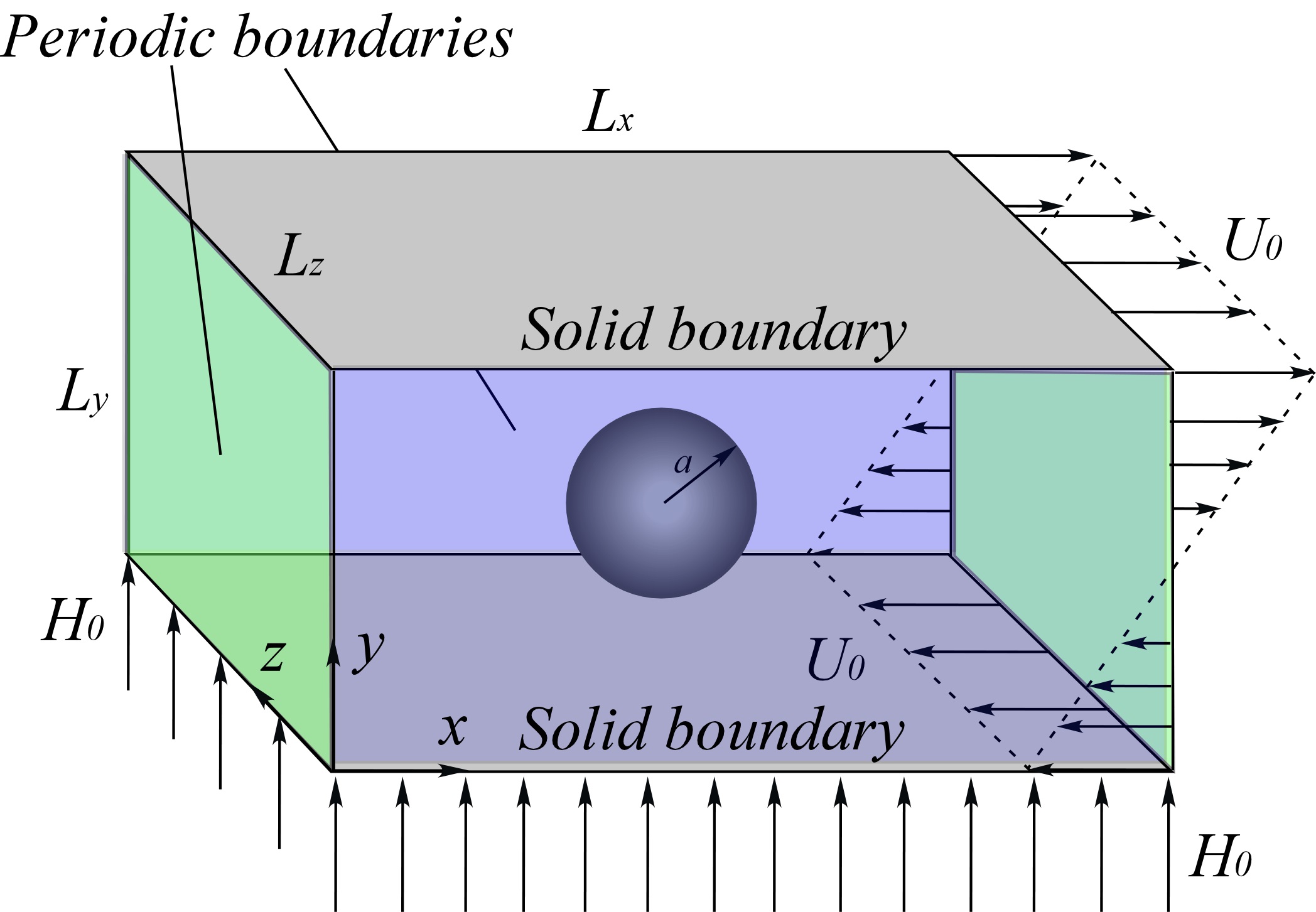}
\caption{\label{fig::schematic_flow}Schematic representation of a drop of ferrofluid inside a Couette cell filled with a non-magnetizable liquid subjected to the simultaneous effects of shear and a wall normal uniform magnetic field of intensity $H_0$.}
\end{figure}
The aim of the present work is to evaluate the role played by magnetic stresses on the rheology of a dilute suspension made of ferrofluid droplets dispersed in a non-magnetic immiscible fluid subjected to the simultaneous effect of a shearing flow and a homogeneous magnetic far field. To accomplish our goal, we consider a Couette cell, as schematized in Fig. \ref{fig::schematic_flow}, consisting of two parallel walls placed at mutual distance $L_y$ {moving in opposite directions with} velocity $\pm U_0\boldsymbol{\mathrm{e}}_x$, and imposing a uniform magnetic field, $\boldsymbol{\mathrm{H}}_0=H_0\boldsymbol{\mathrm{e}}_y$, where $\boldsymbol{\mathrm{e}}_x$ and $\boldsymbol{\mathrm{e}}_y$ are the unit vectors in the $x$- and $y$-axis direction, respectively. 

In order to determine the exact form of the bulk stress which will be used to evaluate the effective viscosity and normal stress difference coefficients, certain assumptions will be made.

Both fluids will be regarded as incompressible and Newtonian, {although, as we have seen} for a ferrofluid the Newtonian constitutive equation for the stresses can not be always safely inferred \textit{a priori} ({further justifications regarding this assumption will be provided below}). Moreover, both phases will be characterized by the same viscosity and density. This latter assumption is necessary to avoid unwanted particle translations {driven by buoyancy}. 

Another assumption which often is tacitly made when dealing with ferrofluids, is the hypothesis that any field-induced non-uniformity of concentration of the ferromagnetic (or ferrimagnetic) particles dispersed in the carrier fluid can be neglected (see \cite{afkhamietal2008, afkhami2010,rowghanian_2016,qiuetal2018,cunha2018,cunha2020}, for instance). This simplification can be questionable, \cite{Stierstadt2015} especially when large magnetic fields are considered (in these conditions, field-induced assemblies may appear even for relatively small particle concentrations \cite{mousavi2015}). Nevertheless, to avoid the difficulty of accounting for fluid density non-homogeneity, and to deal with an expression for the magnetic stress tensor which can be promptly adopted for the calculation of the bulk stress tensor in line with the theory of Batchelor \cite{batchelor_1970},  here we will ignore this complication.

{Additionally}, the concentration of magnetic nanoparticles in the ferrofluid phase is assumed to be sufficiently low that the magnetostatic approximation applies while assuming that the magnetization vector, $\boldsymbol{\mathrm{M}}$, is related to the magnetic field by a linear relation, i.e. we pose $\boldsymbol{\mathrm{M}}=\chi\boldsymbol{\mathrm{H}}$, where $\chi$ is the magnetic susceptibility. This latter assumption restricts considerably the range of applicability of our model, {circumstance that is also shared with the models previously developed by Cunha et al. \cite{cunha2020} and Ishida and Matsunaga \cite{ishida2020}}. In view of these considerations, the following treatment should not be considered complete or general. Nonetheless, it may represent the starting point for ensuing, more accurate, models for the characterization of viscometric functions in the presence of a ferrofluid phase, and provide results at least qualitatively in line with what one should expect in reality (as long as the abovementioned conditions are verified, especially the hypothesis of linearly magnetizable material, arguably the most restrictive).  

{Finally, before embarking on the derivation of the stress model, a final remark regarding the nature of the Maxwell stress tensor (MST), which constitutes the foundation of our derivation: Broadly speaking, a body density force, $\boldsymbol{f}^M$, is said to be Maxwellian if it can be expressed through the divergence of a dyadic field, $\boldsymbol{\nabla}\cdot\boldsymbol{T}^M\equiv\boldsymbol{\nabla}\cdot\boldsymbol{T}^M\left( \boldsymbol{{x}}\right)$, i.e., $\boldsymbol{f}^M= \boldsymbol{\nabla}\cdot\boldsymbol{T}^M$ (see, e.g., \cite{rinaldi_brenner2002}). This definition is rather general and goes beyond the context of electromagnetism; an example being the gravitational density force, $\textbf{\textsl{g}}$, which can be shown to be expressible through the divergence of an adequate gravitational stress field \cite{rinaldi_brenner2002}. In this work, we are dealing with a non-conducting, linearly magnetizable ferrofluid, and the relevant Maxwellian body force can be represented through the divergence of a properly defined Maxwell stress tensor, introduced in the following section. By virtue of this definition, the magnetic density body force can be taken into account by incorporating the Maxwell stress tensor into the true hydrodynamic stress. The theoretical implications of such modus operandi, however, are not as straightforward as one would imagine. Rinaldi and Brenner \cite{rinaldi_brenner2002}, in fact, have pointed out that such operation should be regarded conceptually flawed on a physical ground and may bring to erroneous results in some circumstances which, however, are not a cause of concern in this work. Indeed, for ferrofluid flows, Rinaldi and Brenner \cite{rinaldi_brenner2002} could show that the replacement of the magnetic density force by the corresponding MST counterpart in the linear (and possibly angular) momentum equations, provides correct estimates of the total force (and possibly torque) acting on the fluid domain. On the contrary, the same approach might lead to an erroneous evaluation of the rate of work associated with the Maxwell stress tensor. In the present context, however, the stress model is not affected by the aforementioned limitations since, as we shall see, the MST contribution to the particle extra-stress is ultimately incorporated through the first moment of the magnetic body density force.}

\section{\label{bulk_stress}Bulk stress and rheological properties of the suspension}

As stated before, the main goal of the present work is to investigate the bulk, or effective stress in a dilute suspension of ferrofluid droplets embedded in a non-magnetizable carrier fluid. To accomplish this, we rely on the definition of bulk stress introduced by Batchelor \cite{batchelor_1970}, considered here in its most general formulation. This will allow us to derive a model stress appropriate in the present flow configuration, i.e., in the presence of additional magnetic stresses. One of the key ingredients on deriving the effective stress is the assumption that the two fluids behave as Newtonian, thus before we proceed further we should clarify some aspects related to the rheological properties of ferrofluids.

In the introduction, {we stated that} a ferrofluid on its own may show a variety of rheological features. {In particular, magnetoviscous effects may appear even upon the hypothesis of ideal ferrofluids, i.e., when aggregate formation is not taken into account. This phenomenon, which strictly speaking should not be regarded as a non-Newtonian effect, does not put any particular restriction on the applicability of the model reported in Batchelor \cite{batchelor_1970}, since the viscosity of the ferrofluid phase would be fixed once the extent of the magnetic field is also fixed. Hence, in this regard, care should only be exercised on determining the viscosity of the fluid any time the magnetic field is adjusted. Put more simply, for a given set of experiments performed for a given magnetic field intensity, the viscosity of the ferrofluid remains constant, which is a necessary requirement for the adoption of the model of Batchelor \cite{batchelor_1970}. On the contrary, in the presence of particle aggregation, we have seen that these fluids usually show non-Newtonian responses. Including these effects into the stress model would require the knowledge of reliable constitutive equations for the ferrofluid phase and substantial modifications of the method detailed by Batchelor \cite{batchelor_1970} which are beyond the scope of the present preliminary analysis. On the basis of these considerations, we shall treat the magnetic phase as a Newtonian fluid having a constant viscosity $\eta$, having in mind that possible large discrepancies between experiments and theoretical predictions should primarily be sought among those non-Newtonian features that have been disregarded from the present stress model formulation.}

Another important aspect of the theoretical development detailed by Batchelor \cite{batchelor_1970} that is worth highlighting, is the hypothesis that the resultant of any type of force that might act on the particle should be zero (while allowing for the presence of couples). This hypothesis is required for a definition of the bulk stress that is invariant to translation of the coordinate system. In the following we will see that for a spatially uniform magnetic field there is no net magnetic force acting on the surface of the drop, thus magnetic stresses will be responsible for interface deformations but will not induce drop translations.

Now that the specific requirements necessary for the deduction of the bulk stress in our conditions have been pointed out, we can proceed further with the actual derivation of the model.

Without introducing any restriction on the nature of the particles that may be dispersed in the ambient fluid (e.g., they might be solid particles, drops, capsules, etc.), Batchelor \cite{batchelor_1970} showed that the bulk stress in a suspension is given by the sum of different contributions attributable to the ambient fluid alone and an additional term arising from the presence of the particles. Thus, if we denote with $V$ the whole control volume, and with $V_0$ the volume of a particle of surface area $S_0$, the expression for the bulk stress for a single particle may be written as 
\begin{equation}
\label{eq::bulkstress_general}
\Sigma_{ij} = \frac{1}{V}\int_{V_{a}} -p\delta_{ij} dV  + \eta\left( \frac{\partial U_i}{\partial x_j}+\frac{\partial U_j}{\partial x_i} \right) + \Sigma_{ij}^p,  
\end{equation}
where $V_{a} = V-V_0$ is the volume occupied by the ambient fluid. The remaining variables appearing in Eq. (\ref{eq::bulkstress_general}) are the pressure $p$, the volume-averaged velocity gradient, ${\partial U_i}/{\partial x_j}$, i.e. the average value taken over the whole control volume, being $U_i$ the mean velocity of the imposed flow, differing from the local velocity, $u_i$, arising from the presence of the particles (their difference, $u_i'=u_i-U_i$, can be interpreted as a `perturbation' velocity), and $\delta_{ij}$ is the Kronecker delta. It should be emphasized that, in line with the convention adopted by Batchelor \cite{batchelor_1970},  $S_0$ is defined in such a way it lies on the outside of the interfacial layer, i.e. $V_{a}$ is supposed to be entirely occupied by the ambient fluid. The last term on the right-hand side of Eq. (\ref{eq::bulkstress_general}) represents the extra-stress tensor arising from the presence of the particles, which for negligible inertia may be written as (cf. Batchelor \cite{batchelor_1970})
\begin{equation}
\label{eq::particle_stress_general}
\Sigma_{ij}^p = \frac{1}{V} \int_{V_0} \mathrm{T}_{ij}dV - \frac{1}{V} \int_{S_0} \eta \left(u_in_j + u_jn_i\right)dS,  
\end{equation}
where $\boldsymbol{\mathrm{n}}$ is the unit normal pointing outward the surface $S_0$. Hence, we see that in Stokes flow conditions the contribution to the stress due to the presence of the particles is given by the sum of the volume average (bulk) of the stress tensor $\boldsymbol{\mathrm{T}}$ acting within the particle, and a viscous contribution exerted on its surface by the surrounding fluid. In the particular case of drops having the same viscosity of the ambient fluid, this latter term is {uninfluential}, nevertheless it will be retained for the sake of completeness.

Now, we notice that we may pose (cf. equation (4.3) in Batchelor \cite{batchelor_1970})
\begin{equation}
\label{eq::particle_stress_general_reduced}
\int_{V_0} \mathrm{T}_{ij}dV = \int_{S_0} \mathrm{T}_{ik}x_jn_k dS - \int_{V_0}\frac{\partial \mathrm{T}_{ik}}{\partial x_k}x_j dV,
\end{equation}
thus, this extra-stress can be seen as the sum of a stress acting on the particle surface ({obtained upon the adoption of the divergence theorem}) and a volume integral contribution.

For a ferrofluid drop subjected to a {uniform} magnetic field, the second order tensor $\boldsymbol{\mathrm{T}}$ accounts for two contributions, namely the surface tension stress, $\boldsymbol{\Gamma}$, and the magnetic stress $\boldsymbol{\tau}$. It can been shown, (see, for instance Ref. \cite{batchelor_1970} or Ref. \cite{KENNEDY1994251}) that the surface contribution to Eq. (\ref{eq::particle_stress_general_reduced}) due to surface tension reads 
\begin{equation}
\label{eq::particle_stress_surtens}
\int_{V_0} \mathrm{T}_{ij}dV = \int_{S_0} \sigma k x_jn_k dS
\end{equation}
where $\sigma$ is the surface tension coefficient, and $k$ is the sum of the curvatures of any two orthogonal
sections of the interface containing the local normal $\boldsymbol{\mathrm{n}}$. The second integral of (\ref{eq::particle_stress_general_reduced}) can be shown to be identically zero in this particular case (see, e.g., Batchelor \cite{batchelor_1970}). 

Before we proceed further, some additional observations are required. In the original treatment, Batchelor \cite{batchelor_1970} assumed that any body force per unit volume that might act on the flow was uniformly distributed. Owing to this hypothesis, he concluded that the second term { appearing in the right-hand side} of Eq. (\ref{eq::particle_stress_general_reduced}) may, in general, be equilibrated by {a linearly varying isotropic stress which may be ignored. Hence, with allowance for inertial forces, he pointed out that this term can be replaced by an inertial contribution, $\partial \mathrm{T}_{ij}/\partial x_j=\rho f_i$, where $\rho$ is the fluid density, assumed to be uniform throughout the whole suspension, and $f_i$ is the local acceleration relative to the average value of the acceleration.} Thus, upon the assumption of Stokes flow conditions, he considered this contribution to be negligible. In the present conditions, however, the magnetic body force arises at the interface in the form of a discontinuity, hence the assumption of uniform body force fails and both terms of Eq. (\ref{eq::particle_stress_general_reduced}) should, in principle, be retained. We shall see that the contribution to the bulk stress due to magnetic effects indeed arises from the {volume} integral {in the right-hand side} of Eq. (\ref{eq::particle_stress_general_reduced}). 

\subsection{Extra-stress tensor for ferrofluid droplets under the effect of a homogeneous magnetic field}
\label{sec::particle_extra-stress_tensor}
For a dilute ferrofluid, the magnetostatic approximation may be invoked (e.g., see Refs. \cite{rosensweig2013ferrohydrodynamics, afkhami2010,rowghanian_2016}), thus the magnetic field $\boldsymbol{\mathrm{H}}$ and the magnetic induction  $\boldsymbol{\mathrm{B}}$ are governed by the magnetostatic Maxwell equations
\begin{equation}
\label{eq::magnetostatic_maxwell}
    \boldsymbol{\nabla} \times \boldsymbol{\mathrm{H}} = \boldsymbol{0},\quad \boldsymbol{\nabla} \cdot \boldsymbol{\mathrm{B}} = \boldsymbol{0},
\end{equation}
in which  $\boldsymbol{\mathrm{B}}=\mu_0\mu_r\boldsymbol{\mathrm{H}}$, $\mu_0$ is the magnetic permeability in vacuum and $\mu_r$ is the relative magnetic permeability {of the medium}. For a linearly magnetizable medium, $\mu_r=1+\chi$ while for a non-magnetizable material $\mu_r=1$, since $\chi=0$ in this case.

With these premises, the magnetic stress tensor (MST) in the case of incompressible fluids {can be written as} \cite{rosensweig2013ferrohydrodynamics} 
{\begin{equation}
\label{eq::mag_stress_tensor}
    \boldsymbol{\tau} = -\frac{1}{2} \mu_0 |\boldsymbol{\mathrm{H}}|^2\boldsymbol{\mathrm{I}} + \mu_0\mu_r \boldsymbol{\mathrm{H}} \boldsymbol{\mathrm{H}},
\end{equation}}
i.e., the magnetic stress is the sum of an isotropic term (proportional to the unit tensor $\boldsymbol{\mathrm{I}}$ and to the square of the intensity $H$ of the magnetic field) and a deviatoric part. 

Now, we observe that the drop interface acts as a discontinuity for the MST since the surrounding medium is supposed to be non-magnetizable (i.e., $\mu_r=1$), while inside the ferrofluid phase we have $\mu_r=1+\chi$. A possible way to deal with such a discontinuity is to introduce an indicatrix (cf. Ref. \cite{MELLEMA1983286})
\begin{equation}
\label{eq:B=indictrix}
      \alpha(\boldsymbol{\mathrm{r}})= 
\begin{dcases}
    0,\quad \boldsymbol{\mathrm{r}} \,\,\mathrm{in\, the\, ambient\, fluid}\\
    1,\quad \boldsymbol{\mathrm{r}} \,\,\mathrm{inside\, the\, drop}
\end{dcases}  
\end{equation}
where $\boldsymbol{\mathrm{r}}$ is the vector position, the interface location being identified by the vector $\boldsymbol{\mathrm{r}}_0$. Thus, indicating with the superscript ``\textit{p}'' and ``\textit{a}'' quantities related to the drop (particle) and to the ambient fluid, respectively, the MST reads
\begin{equation}
    \tau_{ij}=\alpha\tau_{ij}^{(p)}+\left( 1-\alpha \right)\tau_{ij}^{(a)},
\end{equation}
therefore, {the magnetic body density force reads}
\begin{equation}
\label{eq::divergence}
    f_i^m=\frac{\partial \tau_{ij}}{\partial x_j} = \alpha\frac{\partial \tau_{ij}^{(p)}}{\partial x_j} + \left( 1-\alpha \right)\frac{\partial \tau_{ij}^{(a)}}{\partial x_j} + \left[\tau_{ij}^{(a)} - \tau_{ij}^{(p)} \right]n_j\delta\left(\boldsymbol{\mathrm{r}} - \boldsymbol{\mathrm{r}}_0 \right),
\end{equation}
where $\delta\left(\boldsymbol{\mathrm{r}} - \boldsymbol{\mathrm{r}}_0 \right)$ is the Dirac delta function. Noting that $\boldsymbol{\nabla} \cdot \boldsymbol{\tau}=\boldsymbol{0}$ everywhere except at the interface location (this is because inside the drop the magnetic particles impose a uniform magnetic field and uniform magnetization, while in the ambient fluid the stress tensor is divergence-free because of the irrotational character of the magnetic field, see, for instance Rowghanian et al. \cite{rowghanian_2016} for further explanations), we find
\begin{equation}
    \frac{\partial \tau_{ij}}{\partial x_j} = \left[\tau_{ij}^{(a)} - \tau_{ij}^{(p)} \right]n_j\delta\left(\boldsymbol{\mathrm{r}} - \boldsymbol{\mathrm{r}}_0 \right).
\end{equation}

Observing that  {$\boldsymbol{\tau}^{(a)}=-\frac{1}{2}\mu_0H^2\boldsymbol{\mathrm{I}}+ \mu_0\boldsymbol{\mathrm{H}}\boldsymbol{\mathrm{H}}$, and $\boldsymbol{\tau}^{(p)}=-\frac{1}{2}\mu_0H^2\boldsymbol{\mathrm{I}}+\mu_0\left(1+\chi \right)\boldsymbol{\mathrm{H}}\boldsymbol{\mathrm{H}}$}, we obtain 
\begin{equation}
     \int_{V_0} \frac{\partial \tau_{ik}}{\partial x_k}x_j dV= -\int_{V_0} \mu_0\chi H_iH_k x_jn_k \delta\left(\boldsymbol{\mathrm{r}} - \boldsymbol{\mathrm{r}}_0 \right)dV =-\int_{S_0} \mu_0\chi H_iH_kx_jn_k dS.   
\end{equation}

{Evaluation of the role played by the} remaining term, i.e., the surface integral of Eq. (\ref{eq::particle_stress_general_reduced}), {requires special considerations. First of all, we observe that the magnetic force will be introduced into the momentum equation (shown in the subsequent section) through the divergence of the Maxwell stress tensor. Thus, with regard to magnetic effects, the contribution to the exchange of momentum is provided by the magnetic body force, $\boldsymbol{f}^m=\boldsymbol{\nabla} \cdot \boldsymbol{\tau}$. In view of this, we conclude that hydrodynamic stresses of magnetic nature are generated by the sole force $\boldsymbol{f}^m$ and the surface contribution to the particle extra-stress should not be accounted to what concerns magnetic effects.} 

{On the basis of the previous considerations, the particle stress tensor finally reads}
\begin{equation}
\label{eq::stresslet_final}
\Sigma_{ij}^p = \frac{1}{V} \int_{S_0}  \{ \sigma k x_jn_i  -\eta\left(1-\lambda \right) \left(u_in_j + u_jn_i\right) + \mu_0\chi H_iH_kx_jn_k \}  dS,
\end{equation}
where $\lambda=\eta_p/\eta$ is the ratio between the viscosities of the droplet and of the ambient fluid. Since in the present work $\lambda=1$, this term {will not be taken into account}. {We may note that this formulation is rather general and could also be applied to an ``inverse'' emulsion (i.e., for non-magnetizable drops surrounded by a ferrofluid) or when both phases are magnetizable. In this regard, assuming the surrounding phase and drop characterised by magnetic susceptibilities, $\chi^{a}$ and $\chi^{p}$, respectively, it would be sufficient to use the term $\mu_0\left( \chi^{p}-\chi^{a}\right)$ in place of $\mu_0\chi$.}

{It should be emphasized that the stress Eq. (\ref{eq::stresslet_final}) is meaningful only for a zero-thickness interface. As we shall see, in the framework of the numerical approach adopted here the drop boundary is represented by a finite thickness layer in which $\alpha$ is a continuously varying function. Thus, within the interfacial region the two divergence terms previously disregarded from Eq. (\ref{eq::divergence}) are not identically vanishing functions and should be re-introduced in the numerical implementation of the model. We will come back to this aspect later, when we will describe the approach in the context of the numerical framework.}

{We might also note that the presence of a magnetic force at the drop-fluid interface generates a torque, thus, contrarily to the surface tension tensor, the magnetic particle stress tensor is not symmetric in general. Hence, for some purposes, it might be convenient separating it in its symmetric and antisymmetric parts. With the obvious meaning of the symbols adopted (the superscript $m$ indicates that we are considering only the magnetic term of Eq.(\ref{eq::stresslet_final})), we have, 
\begin{equation}
    S_{ij}^m= \frac{1}{2}\left( \Sigma_{ij}^{p,m} + \Sigma_{ji}^{p,m}\right), \quad A_{ij}^m= \frac{1}{2}\left( \Sigma_{ij}^{p,m} - \Sigma_{ji}^{p,m}\right).
\end{equation}
The antisymmetric part of the magnetic particle extra stress is related to the magnetic torque, $\boldsymbol{\mathrm{C}}^m$, through the simple relationship $A_{ij}^m = -\frac{1}{2}\varepsilon_{ijk}C_k^m$, where $\varepsilon_{ijk}$ is the Levi-Civita symbol.}

Once the {particle} stress (\ref{eq::stresslet_final}) has been computed, the steady shear rheology, which is characterized by the ``excess'' viscosity, $\eta_e$, and the two normal stress differences normalised by the reference shear stress $\eta \dot{\gamma}$ (in the following, $N_1$ and $N_2$ are simply termed as dimensionless normal stress differences for the sake of brevity), is given by
\begin{equation}
\label{eq::rheological_functions}
  \begin{split}
      \frac{\eta_e}{\eta}=1+\frac{\Sigma_{xy}^p}{\eta \dot{\gamma}},\quad
    {N_1} = \frac{\Sigma_{xx}^p-\Sigma_{yy}^p}{\eta \dot{\gamma}}, \quad
    {N_2} = \frac{\Sigma_{yy}^p-\Sigma_{zz}^p}{\eta \dot{\gamma}},
\end{split}  
\end{equation}
where $\dot{\gamma}$ is the imposed shear rate.

Finally, we may define an average shear stress evaluated as (see, for instance \cite{cunha2020})
\begin{equation}
\label{eq::excess_stress_integral}
    \Sigma_{xy} = \frac{1}{S_w}\int_{S_w}\eta \frac{\partial u_x}{\partial y}dS,
\end{equation}
where $S_w$ represents indistinctly the surface of one of the two lateral walls $y=0$ or $y=1$.  {From this expression, and with the aid of Eq. (\ref{eq::bulkstress_general}), we can finally work out an alternative expression for the particle extra shear stress, 
\begin{equation}
\label{eq::excess_shear_stress}
    \Sigma_{xy}^p = \Sigma_{xy}-\dot{\gamma}\eta.
\end{equation}}

\section{\label{math_num_models}Mathematical and numerical models}

The set of governing equations is solved numerically in a Cartesian frame of reference using a hybrid level set-volume-of-fluid based OpenFOAM code developed by Capobianchi et al. \cite{CAPOBIANCHI2018313} {on the basis of the original formulation of Yamamoto et al. \cite{takuya2017}}. Here, we highlight the general features of the methodology, while the reader is addressed to Capobianchi et al. \cite{CAPOBIANCHI2018313} for a detailed description of the approach.

Firstly, we observe that the discrete counterpart of the magnetostatic Maxwell equations (\ref{eq::magnetostatic_maxwell}) may be rewritten in terms of a scalar potential $\psi$ in the following manner
\begin{equation}
\label{eq::magnetostatic_maxwell_discrete}
 \boldsymbol{\mathrm{H}} =-\boldsymbol{\nabla}\psi,\quad  \boldsymbol{\nabla}\cdot \left( \mu(\boldsymbol{\mathrm{x}}) \boldsymbol{\nabla}\psi\right)=0.
\end{equation}
On writing the second equation of (\ref{eq::magnetostatic_maxwell_discrete}), we adopted the ``one-fluid'' formulation, having highlighted the fact that in this context the magnetic permeability is regarded as a continuous quantity  $\mu(\boldsymbol{\mathrm{x}})=\alpha(\boldsymbol{\mathrm{x}})\left(1+\chi \right)\mu_0 + \left(1-\alpha(\boldsymbol{\mathrm{x}}) \right)\mu_0$ through the discrete vector position, $\boldsymbol{\mathrm{x}}$. Here, $\alpha(\boldsymbol{\mathrm{x}})$ is the standard fraction function adopted in VOF-based codes, which can be regarded as the discrete counterpart of the indicatrix function $\alpha(\boldsymbol{\mathrm{r}})$ introduced in Sect.\ref{sec::particle_extra-stress_tensor}. Generally speaking, similar definitions apply for any other material property that may be encountered in the problem. Since we are dealing with isodense and isoviscous fluids, density and viscosity are constant in space, however, on writing the governing equations, the functional dependence of these two quantities on the position $\boldsymbol{\mathrm{x}}$ will be retained for the sake of generality.

The fluid flow obeys the isothermal and incompressible conservation of mass and Navier-Stokes equations for magnetizable fluids in the presence of a magnetic vector field 
\begin{equation}
\label{eq::N-S}
    \boldsymbol{\nabla}\cdot\boldsymbol{\mathrm{u}}=0,\quad   \rho(\boldsymbol{\mathrm{x}})\left({\partial }/{\partial t} + \left(\boldsymbol{\mathrm{u}}\cdot \boldsymbol{\nabla}\right)\right) \boldsymbol{\mathrm{u}}=-\boldsymbol{\nabla} p+\boldsymbol{\nabla}\cdot\left( 2\eta(\boldsymbol{\mathrm{x}})\boldsymbol{\mathrm{D}} \right)+\boldsymbol{\mathrm{f}}^{\sigma}+\boldsymbol{\mathrm{f}}^m,
\end{equation}
where $\rho(\boldsymbol{\mathrm{x}})$ is the density and $\boldsymbol{\mathrm{D}}=\frac{1}{2}\left(\boldsymbol{\nabla}\boldsymbol{\mathrm{u}}+ \left(\boldsymbol{\nabla}\boldsymbol{\mathrm{u}} \right)\right)^{\mathrm{T}}$ is the rate-of-strain tensor. The two forces {densities} appearing on the right-hand side of the momentum equation (\ref{eq::N-S}) account for the surface tension and the magnetic force. {The former can be written as $\boldsymbol{\mathrm{f}}^{\sigma} = \sigma k\left(\varphi \right)\boldsymbol{\mathrm{n}}\left(\varphi \right) \delta\left(\varphi \right)$ (cf. Ref. \cite{BRACKBILL1992335}) where $\boldsymbol{\mathrm{n}}\left( \varphi\right)=-\frac{\boldsymbol{\nabla}\varphi}{||\boldsymbol{\nabla}\varphi||}$ is the (discrete) outward normal at the drop interface, $k\left(\varphi \right)=\boldsymbol{\nabla}\cdot\boldsymbol{\mathrm{n}}\left(\varphi \right)$ and $\varphi$ is the level set function (see, e.g., Refs. \cite{CAPOBIANCHI2018313,takuya2017} for more information).} The {magnetic body force density}, on the basis of the assumptions made, reads as
{\begin{equation}
    \boldsymbol{\mathrm{f}}^m = \boldsymbol{\nabla}  \cdot \left[ - \frac{1}{2}\mu_0 {{\left| \boldsymbol{\mathrm{H}} \right|}^2}\boldsymbol{\mathrm{I}} + {\mu\left(\boldsymbol{\mathrm{x}} \right) \boldsymbol{\mathrm{H}} \boldsymbol{\mathrm{H}} } \right].
\end{equation}}
As anticipated, this force vanishes everywhere apart from at the interface since in the bulk of each phase the divergence of the magnetic stress tensor is identically zero. {In the present numerical framework the interface is characterized by a finite thickness, within which the MST is not divergence-free ($0<\alpha<1$). This fact must be taken into account on evaluating the magnetic part of the extra-stress tensor. Reintroducing the divergence terms discharged from Eq. (\ref{eq::divergence}), the magnetic part of the bulk stress now assumes the compact form
\begin{equation}
\label{eq::model_numerical}
    \boldsymbol{\Sigma}^{p,m}=\frac{1}{V}\int_V  -\boldsymbol{\mathrm{f}}^m \otimes \boldsymbol{\mathrm{x}} dV.
\end{equation}
Note that the domain of integration can be conveniently extended to the entire domain since, for the reasons explained before, the magnetic force is zero everywhere except at the interface.}

{With reference to Eq. (\ref{eq::model_numerical}), we observe that if the origin of the coordinate system is shifted by an arbitrary vector $\boldsymbol{\mathrm{x}}_0$, we have
\begin{equation}
\label{eq::model_numerical_translated}
    \boldsymbol{\Sigma}^{p,m}\left( \boldsymbol{\mathrm{x}} - \boldsymbol{\mathrm{x}}_0\right)=-\frac{1}{V}\int_V  \boldsymbol{\mathrm{f}}^m \otimes \left( \boldsymbol{\mathrm{x}} - \boldsymbol{\mathrm{x}}_0\right) dV = -\frac{1}{V}\int_V  \boldsymbol{\mathrm{f}}^m \otimes \boldsymbol{\mathrm{x}} dV + \frac{1}{V}\left\{\int_V  \boldsymbol{\mathrm{f}}^m dV \right\} \otimes \boldsymbol{\mathrm{x}}_0,
\end{equation}
but previously we have anticipated that the rightmost integral of Eq. (\ref{eq::model_numerical_translated}) must vanish for uniform magnetic fields, hence the statement made regarding the arbitrariness of the origin of the coordinate axes mentioned at the beginning of this section follows consequently.}

{Regarding the surface tension extra-stress tensor, we observe that a similar approach could be used. Indeed, in the present numerical framework, the surface tension contribution could be accounted for with an additional surface tension density force, $\boldsymbol{\mathrm{f}}^{\sigma}$, added within the volume integral of Eq. (\ref{eq::model_numerical}), as done by Ishida and Matsunaga \cite{ishida2020}. Nevertheless, we also note that this method is not strictly required, since the interfacial tension is a constant, while the remaining variables are purely geometrical quantities (cf. Eq. (\ref{eq::stresslet_final})), meaning that the variable $\alpha$ is not involved here. Hence, once the interface location has been identified (iso-surface $\alpha=0.5$), the interfacial extra-stress can be calculated through the aid of Eq. (\ref{eq::stresslet_final}). Practically speaking, this operation was accomplished in postprocessing by extracting the surface and calculating the integral
\begin{equation}
    \Sigma_{ij}^{{p,\sigma}}=\int_{S_0} \{\sigma\left(\delta_{ij}-n_in_j\right)\} dS,
\end{equation}}
{taking advantage of the identity $kx_jn_i=\delta_{ij}-n_in_j$ (cf. Refs. \cite{batchelor_1970,li_sarkar2005}). This approach was found to be numerically more accurate, because the normal vector computed from the reconstructed interface was found to be generally more precise than the one evaluated by computing the gradient of the level-set function (this latter quantity, in turn, would serve to compute the force $\boldsymbol{\mathrm{f}}^{\sigma}$).}

The governing equations (\ref{eq::magnetostatic_maxwell_discrete}) and (\ref{eq::N-S}) are discretized in a three-dimensional computational domain having dimensions $\left(L_x=2, L_y=1, L_z=1 \right)$ composed of $\left(120 \times 60 \times 60 \right)$ cells in the respective directions, $x$, $y$ and $z$ (mesh $M_0$). An initially spherical drop of radius $a=0.1$ is placed at the centre of the computational Couette cell, i.e., its centre being placed at the point of coordinates $\left(1, 1/2, 1/2 \right)$. An octree adaptive mesh refinement is employed at the interface, adopting three consecutive levels of refinement within an iteration (the typical refined cell at the interface is cube having sides $2^3$ times smaller than the parent (non-refined) cell), with the refined mesh consisting of about 1.5M nodes. At the boundaries $y=0$ and $y=1$, Dirichlet boundary conditions are applied for the velocity by imposing $\boldsymbol{\mathrm{U}}=\left(-U_0,0,0\right)$ and $\boldsymbol{\mathrm{U}}=\left(U_0,0,0\right)$ respectively, yielding to a constant shear rate $\dot{\gamma}=2U_0/L_y\equiv2U_0$. A uniform magnetic far field vector, $\boldsymbol{\mathrm{H}}=\left( 0,H_0,0 \right)$, is set by imposing the conditions $\psi=\psi_0$ and $\psi=\psi_1$ at the boundaries $y=0$ and $y=1$, respectively so that the resulting magnetic field is $H_0=\left(\psi_1-\psi_0\right)/L_y\equiv\psi_1-\psi_0$. Periodic flow conditions are applied at the remaining boundaries, i.e., at $x=0$ and $x=2$, and at $z=0$ and $z=1$.

Since we are considering periodic conditions, hydrodynamic interactions between two adjacent droplets may come into  play due to the relatively short extension of the domain. Confinement effects in the $y$-direction may also be relevant, especially for those cases where the relative strength of viscous and magnetic effects are predominant with respect to the interfacial tension (we shall see later that in these cases the drop appears largely stretched and partially aligned to the magnetic field, thus the relative distance between the poles of the drop and the lateral wall can be critically small). The role played by these effects {has been evaluated considering a droplet with halved radius, maintaining the domain size and mesh resolution, for flow conditions that provided the largest drop elongation in the vertical direction. No substantial differences were observed in relation to the corresponding case for the original drop radius, therefore the effect of confinement can be regarded negligible for the conditions considered here. Moreover, a mesh study was conducted considering the largest value of $\mathrm{Bo_m}$ (i.e., as we shall see, $\mathrm{Bo_m}=5.6$) for three different levels of refinement by halving the original mesh size, $M_0$, one time (mesh $\mathrm{M}_1$), twice (mesh $\mathrm{M}_2$) and finally three times (mesh $\mathrm{M}_3$). A good rate of convergence was found and all subsequent simulations have been carried out using the resolution $\mathrm{M}_3$. Detailed information regarding the confinement and mesh studies can be found in the Supplementary Information document.}
\\
\indent Prior to embarking on the discussion of the results, we list the set of non-dimensional parameters that will be used. Adopting $a$, $\dot{\gamma}a$, $\dot{\gamma}^{-1}$, {$\eta\dot{\gamma}$} and $H_0$ as reference quantities for length, velocity, time, {stress} and magnetic field, respectively, we can define the Reynolds number, $\mathrm{Re}=\rho \dot{\gamma}a^2/\eta$, the capillary number, $\mathrm{Ca}=\eta\dot{\gamma}a/\sigma$ and the magnetic Bond number $\mathrm{Bo_m}=\mu_0 {H_0}^2a/\sigma$. In the present context, $\mathrm{Re}\ll1$, thus inertial effects can be neglected. The remaining two parameters represent the ratio between viscous force and interfacial tension ($\mathrm{Ca}$), and the ratio between the magnetic force and interfacial tension ($\mathrm{Bo_m}$), hence the drop dynamics depend exclusively on the interplay between viscous stresses, interfacial tension and magnetic stresses. 

\section{\label{results}Results}

\subsection{Drop morphology and rheological functions in the absence of magnetic field: comparison with existing theoretical models}
\label{sec::validation}
Before discussing the role of the magnetic stress on the rheological properties of the system in the presence of magnetic effects, we assess the accuracy of the numerical approach assuming $\boldsymbol{\mathrm{H}}_0=\boldsymbol{0}$ by comparing our results against two different theoretical models, namely the model proposed by Choi and Schowalter \cite{choi1975} (C-S model), and the morphological model of Yu et al. \cite{yu_bousmina2002} (GBP-YB model), which was based on the earlier work of Grmela et al. \cite{garmelaetal2001}.
\begin{figure}[h!]
\centering
\includegraphics[width=0.95\textwidth]{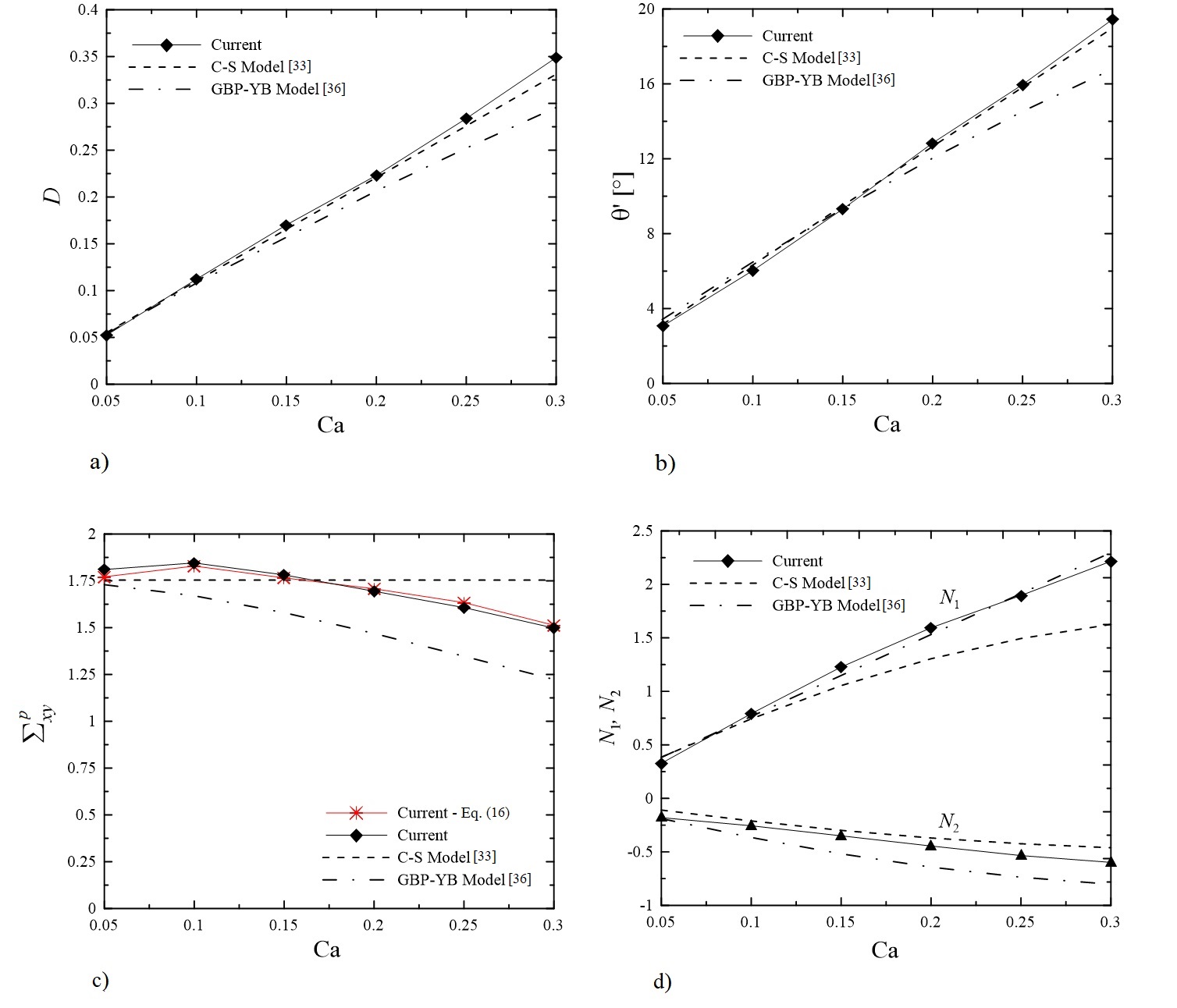}
\caption{\label{fig::validation} Deformation parameter (a), drop orientation (b), excess shear stress (c) and normal stress differences (d) given as functions of the Capillary number. The present results are compared with the C-S model \cite{choi1975} and the GBP-YB model \cite{yu_bousmina2002}. The excess stresses in (c) are scaled with the quantity $\eta\dot{\gamma}\phi$, while normal stresses (d) are scaled with the volume fraction $\phi$.}
\end{figure}

To this end, Fig. \ref{fig::validation}a shows the drop deformation parameter, {$D=\left(a_1-a_2\right)/\left(a_1+a_2\right)$} (here {$a_1$ and $a_2$} are the major and minor axes of the drop measured on the mid-plane $z=0.5$), respectively as a function of $\mathrm{Ca}$ obtained with the present simulations (squared symbols) compared to the two theoretical models mentioned before. For small values of $\mathrm{Ca}$, the numerical simulations and the models predict similar deformations. As $\mathrm{Ca}$ is increased, however, the numerical simulations always provide larger deformations. Analogously, in Fig. \ref{fig::validation}b we show the comparison in terms of droplet orientation, $\theta'=45^{\circ}-\theta $ (measured in degrees), where $\theta$ is the angle between the drop major axis, {$a_1$}, and the $x$-axis, i.e., the axis oriented along the direction of the unperturbed flow (cf. Fig. \ref{fig::def_orient_magnetic}b). It can be seen that the C-S model is in good agreement with our simulations in the whole range of $\mathrm{Ca}$, while the GBP-YB model consistently predicts smaller values of $\theta'$. Comparisons with previous numerical simulations are provided as supplementary {information}.

Figs. \ref{fig::validation}c,d show the comparison in terms of dimensionless excess shear stress, $\Sigma_{xy}^p$, and normal stress differences, $N_1$, $N_2$, respectively (with abuse of notation, unless otherwise stated, in the following we shall indicate normalized stresses with the same symbolism adopted for the respective dimensional quantity, e.g., $\boldsymbol{\mathrm{\Sigma}}^p\equiv \boldsymbol{\mathrm{\Sigma}}^p/\left(\phi\dot{\gamma}\eta\right)$, being customary to divide stresses by the volume fraction $\phi=V_0/V$ of the ferrofluid phase dispersed in the ambient fluid). The latter will be kept constant throughout the whole study and equal to $\phi\simeq 0.21\%$.

In Fig. \ref{fig::validation}c, we observe that the C-S model provides an excess stress that is independent of $\mathrm{Ca}$ (see the corresponding dashed line), contrarily to the GBP-YB  model which correctly reproduces the shear-thinning behavior. Moreover, we note that for very low $\mathrm{Ca}$ both models and the present numerical results (diamond symbols) roughly predict the same excess stress $\Sigma_{xy}^p$ (i.e., the same excess viscosity $\eta_e$). As $\mathrm{Ca}$ increases, the simulations capture the expected shear-thinning behaviour although the simulation data are consistently larger than the values provided by the GBP-YB model. A similar discrepancy was also observed by Li and Sarkar \cite{li_sarkar2005} for a numerical model system analogous to the present one (same values of the domain confinement but larger value of the Reynolds number, $\mathrm{Re}=0.1$).  Additionally, in the same figure we show the excess stress evaluated by means of Eq. (\ref{eq::excess_shear_stress}) (red symbols). The good agreement with the values calculated by means of Eq. (\ref{eq::stresslet_final}) is excellent over the whole range of capillary number considered, which indirectly highlights the reliability of the methodology adopted to evaluate the quantities appearing in expression (\ref{eq::stresslet_final}).

With regard to the normal stress differences, from Fig. \ref{fig::validation}d we observe an excellent agreement between our predictions (diamond symbols) and the GBP-YB model in terms of $N_1$ in the whole range of $\mathrm{Ca}$, while the C-S model provides an underestimation of $N_1$ if compared to the other sets of data. Regarding the second normal stress difference, $N_2$, the present simulations (filled triangles) provide values lying in between those predicted by the two theoretical models.

Overall, we can conclude that the present numerical approach is in qualitative agreement with both the C-S and the GBP-YB model. At large $\mathrm{Ca}$ however, a certain deviation is observed, which is expected since the theoretical models were conceived in the framework of small $\mathrm{Ca}$ theories. 

\subsection{Emulsion rheology in the presence of the magnetic field}
\begin{figure}
\centering
\includegraphics[width=0.95\textwidth]{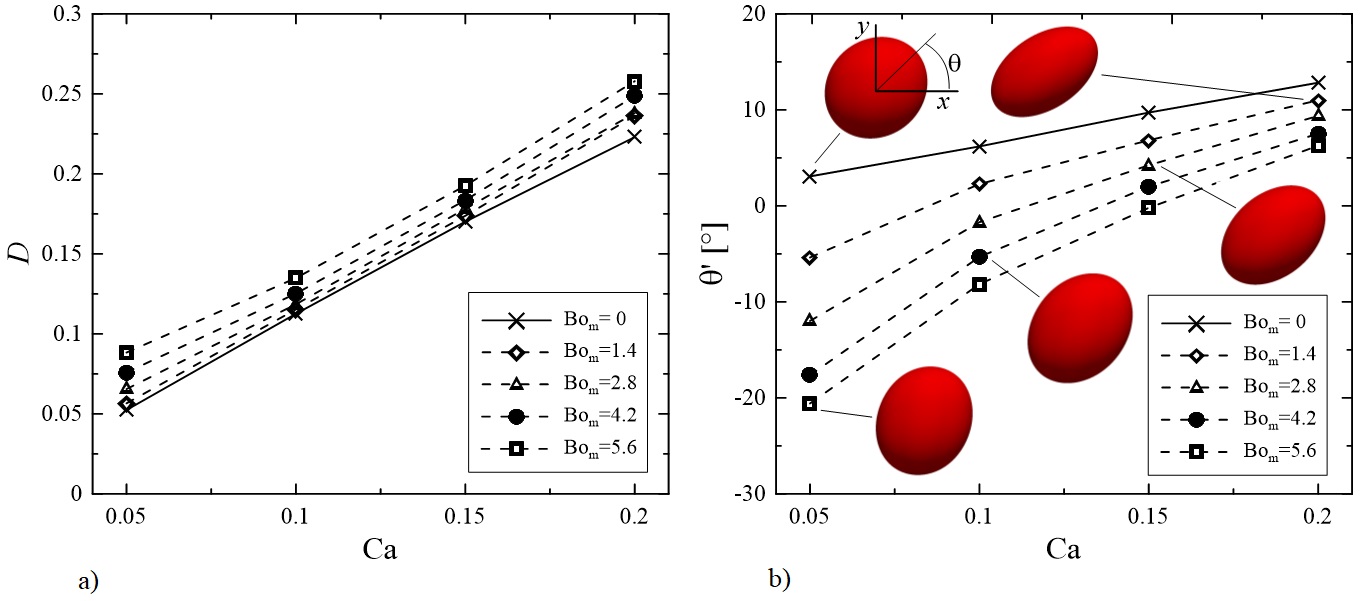}
\caption{\label{fig::def_orient_magnetic} Deformation (a) and orientation (b) as a function of $\mathrm{Ca}$ for different values of the magnetic Bond number, $\mathrm{Bo_m}$. $\chi=0.5$. Some representative images of the drops for different conditions have been added in (b).}
\end{figure}
The accuracy of the ferrofluid solver was already assessed by Capobianchi et al. \cite{CAPOBIANCHI2018313} considering a ferrofluid droplet undergoing deformation under the effect of a spatially uniform magnetic field and absence of flow. The results were compared with the experiment of Afkhami et al. \cite{afkhami2010} and the calculations of Rowghanian et al. \cite{rowghanian_2016} under the assumption of linearly magnetizable material (confirmed by the experimental observations of \cite{afkhami2010}, {whose magnetization curves are shown in Appendix \ref{sect::appendixB}}) and for a value of the magnetic susceptibility $\chi=0.8903$, which was considered during the experiments of Afkhami et al. \cite{afkhami2010} and ensuing calculations of Rowghanian et al. \cite{rowghanian_2016}. In the following, however, we shall consider a smaller susceptibility ($\chi=0.5$) since for large magnetic Bond numbers and $\chi=0.8903$ we observed the presence of an instability which would deserve a separate investigation. The range of magnetic fields considered (i.e., the range of $\mathrm{Bo_m}$ in practice) will be the same as in Capobianchi et al. \cite{CAPOBIANCHI2018313} for which we assume the validity of the magnetic linear constitutive equation. All other material parameters are the same as in the cases of Sect. \ref{sec::validation} {(the interested reader will find in Appendix \ref{sect::appendixB} some considerations about the dimensional values corresponding to these dimensionless quantities). Additionally, it is worth mentioning that for the fluid pair adopted by Afkhami et al. \cite{afkhami2010} in their low-magnetization regime, appreciable drop interface displacements would appear for millimetre-sized drops or moderately smaller. Nevertheless, using different fluid pairs, the interfacial tension can be drastically lowered and drop deformations can be appreciably large even for microsized drops upon the application of moderate magnetic fields \cite{Zakinyan2011}. Considerations regarding the values of drop deformation to be expected in actual experiments using both the parameters reported in Afkhami et al. \cite{afkhami2010} and in Zakinyan and Dikansky \cite{Zakinyan2011} can be found in Appendix \ref{sect::appendixB}.}

Fig. \ref{fig::def_orient_magnetic}(a) shows the deformation, $D$ versus $\mathrm{Ca}$ for different values of the magnetic Bond number, $\mathrm{Bo_m}$. It can be seen, as expected, that for a fixed $\mathrm{Bo_m}$ the deformation increases monotonically with $\mathrm{Ca}$ due to the increased shear stress in the face of a constant interfacial tension. Essentially, the trends are therefore congruent with the behaviour observed for the non-magnetic case shown in Fig. \ref{fig::validation}(a). Analogously, a monotonic increase is observed also for increasing values of $\mathrm{Bo_m}$ for fixed values of $\mathrm{Ca}$, since the magnetic stresses act to ``stretch'' the drop in the direction of the imposed magnetic vector field, thereby contributing to increment the drop surface while forcing it to be oriented toward the vertical axis due to the presence of a magnetic torque. In this regard, Fig. \ref{fig::def_orient_magnetic}(b) shows the corresponding orientation $\theta'$. Again, we note a similar monotonic behaviour with $\mathrm{Ca}$, while $\theta'$ decreases for increasing $\mathrm{Bo_m}$ for each value of $\mathrm{Ca}$. Moreover, it is worthwhile highlighting that for most of the conditions considered here $\theta'$ is negative ($\theta>45^{\circ}$). Only for $\mathrm{Ca}=0.2$ the orientation $\theta'$ is positive for all values of $\mathrm{Bo_m}$ due to increasingly strong viscous effects which act to orient the drop toward the direction of the imposed shear. Some of the shapes obtained for different combinations of $\mathrm{Ca}$ and $\mathrm{Bo_m}$ have been added in the figure for the sake of clarity.

\begin{figure}
\centering
\includegraphics[width=0.95\textwidth]{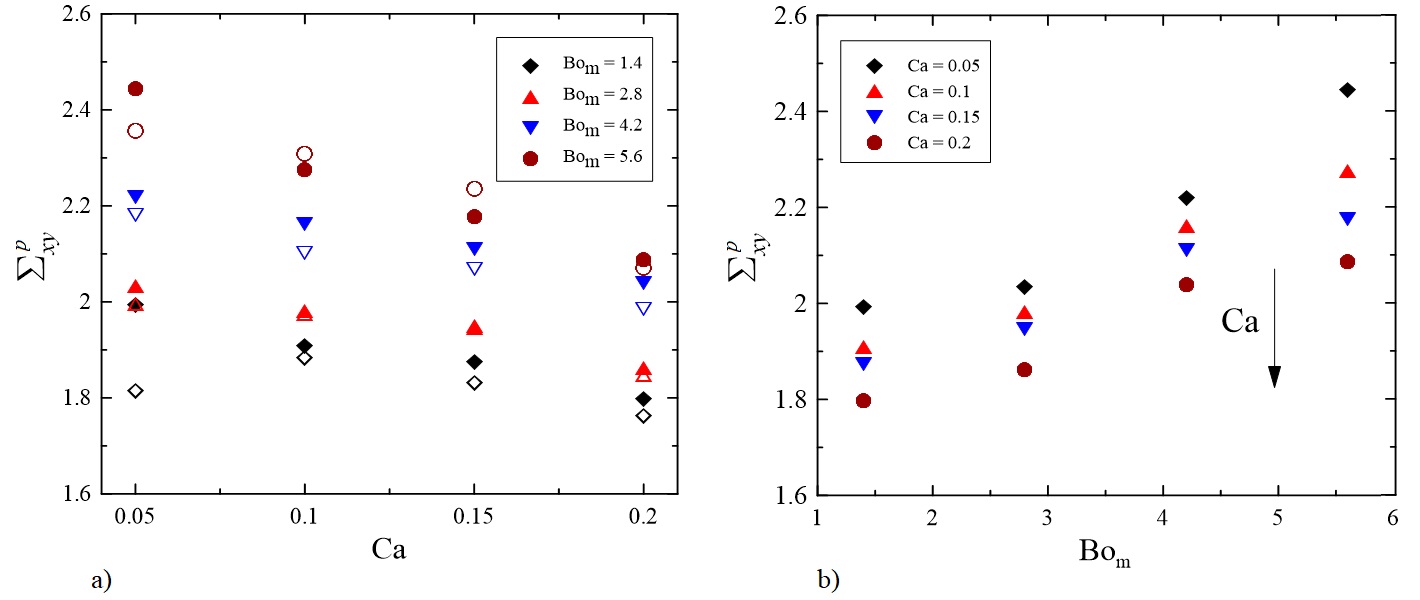}
\caption{\label{fig::sigma_excess}{a) Excess shear stress as a function of $\mathrm{Ca}$ for different $\mathrm{Bo_m}$. b)Excess shear stress as a function of $\mathrm{Bo_m}=$ for different $\mathrm{Ca}$. $\chi=0.5$. Open symbols represent data obtained from Eq. (\ref{eq::stresslet_final}), while data represented with closed symbols were obtained through Eq. (\ref{eq::excess_shear_stress}). Data shown are scaled with the quantity $\eta\dot{\gamma}\phi$.}}
\end{figure}

As a result of the relevant modification of the droplet morphology induced by the additional magnetic stresses, the rheological properties of the emulsion are expected to be substantially different from those observed in the absence of magnetic effects. In this regard, in Fig. \ref{fig::sigma_excess}a we show the particle excess shear stress, $\Sigma_{xy}^p$, as a function of $\mathrm{Ca}$ for different values of the magnetic Bond number obtained by means of the current model {(open symbols)} and compare them with the values obtained by means of Eq. (\ref{eq::excess_shear_stress}) {(closed symbols)}. In general, we notice a shear-thinning behaviour, although the relative variations of $\Sigma_{xy}^p$ with $\mathrm{Ca}$ are less pronounced than those obtained in the absence of magnetic effects (cf. Fig. \ref{fig::validation}c), with the exception of the results obtained with the present model for low $\mathrm{Ca}$ regime, here represented by $\mathrm{Bo_m}=1.4$ (cf. Fig. \ref{fig::sigma_excess}a). Such discrepancy can be attributed to the numerical error associated to the evaluation of the components of the vector normal to the drop interface when the interfacial tension is predominant with respect to the other constraints, which is a typical drawback of interface capturing techniques such as VOF and level-set methods (see, e.g., Refs. \cite{BRACKBILL1992335,MENARD2007510}). Nevertheless, it is worth pointing out that the trend obtained through Eq. (\ref{eq::excess_shear_stress}) is consistent with that obtained through {the adoption of the model} in all the other cases, i.e., it exhibits monotonic decreasing behaviour in the whole range of $\mathrm{Ca}$. 

From Fig. \ref{fig::sigma_excess}b, where we report the excess shear stress obtained by means of Eq. (\ref{eq::excess_shear_stress}) for all values of $\mathrm{Bo_m}$, we can infer that $\Sigma_{xy}^p$ increases for increasing $\mathrm{Bo_m}$ for a fixed $\mathrm{Ca}$. In particular, with reference to Fig. \ref{fig::sigma_excess}b, we notice a monotonic increase (``magneto-thickening''), roughly {cubic} behaviour, suggesting the possibility to model the emulsion viscosity with an equation like 
\begin{equation}
\label{eq::const_equation}
    \frac{\eta_{e}}{\eta}\approx 1+\phi\Sigma_{xy}^{p,(0,0)} + {\phi \left( K_{\chi,1}\mathrm{Bo_m} + K_{\chi,2}\mathrm{Bo_m}^2 + K_{\chi,3}\mathrm{Bo_m}^3 \right)}, 
\end{equation}
which applies for fixed capillary numbers. Here, $\Sigma_{xy}^{p,(0,0)}$ represents the excess stress when $ \mathrm{Bo_m}\to0$, while {$K_{\chi,i}$ ($i=1,2,3$) are constants of proportionality which are} expected to be dependent on the magnetic susceptibility (it appears reasonable, in fact, to expect an increase of $\eta_{e}$ with $\chi$ since larger values of $\chi$ lead to larger magnetic stresses). However, we should recall that the present model is valid only for linearly magnetizable fluids, which limits the maximum value of $\chi$ (see, e.g., Ref. \cite{Stierstadt2015}). Finally, the direct linear proportionality to the volume fraction $\phi$ is a consequence of the additive character of the model (\ref{eq::stresslet_final}) (see e.g., Batchelor \cite{batchelor_1970} and Li and Sarkar \cite{li_sarkar2005} for additional information), and it seems reasonable to assume that such behaviour is valid as far as dilute regimes are concerned.
\begin{figure}[h!]
\centering
\includegraphics[width=0.95\textwidth]{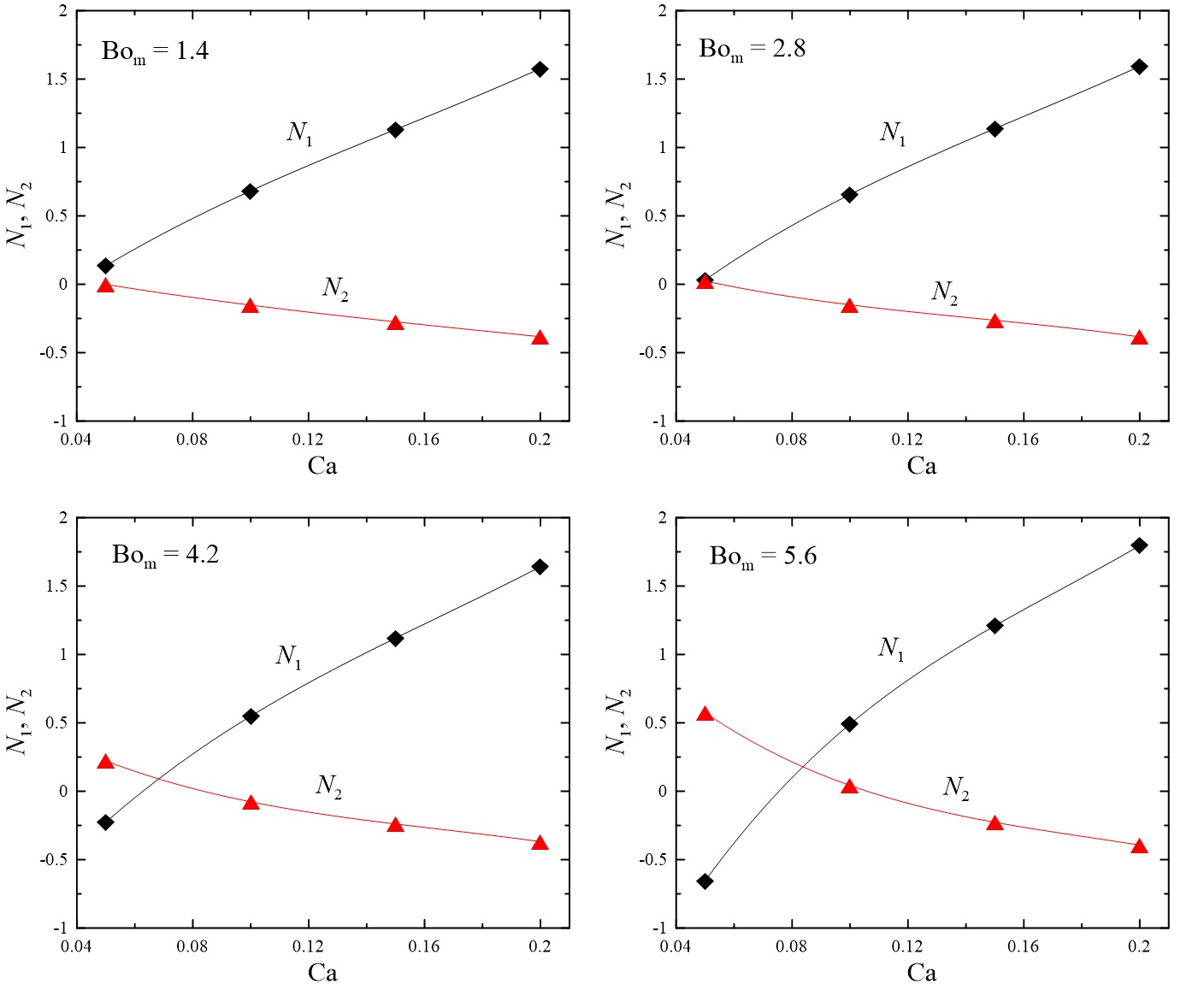}
\caption{\label{fig::normal_stresses} {First and second normal stress differences as a function of $\mathrm{Ca}$ for different $\mathrm{Bo_m}$. $\chi=0.5$. The lines represent cubic fits. Data are scaled with the volume fraction $\phi$.}}
\end{figure}
\begin{figure}
\centering
\includegraphics[width=0.95\textwidth]{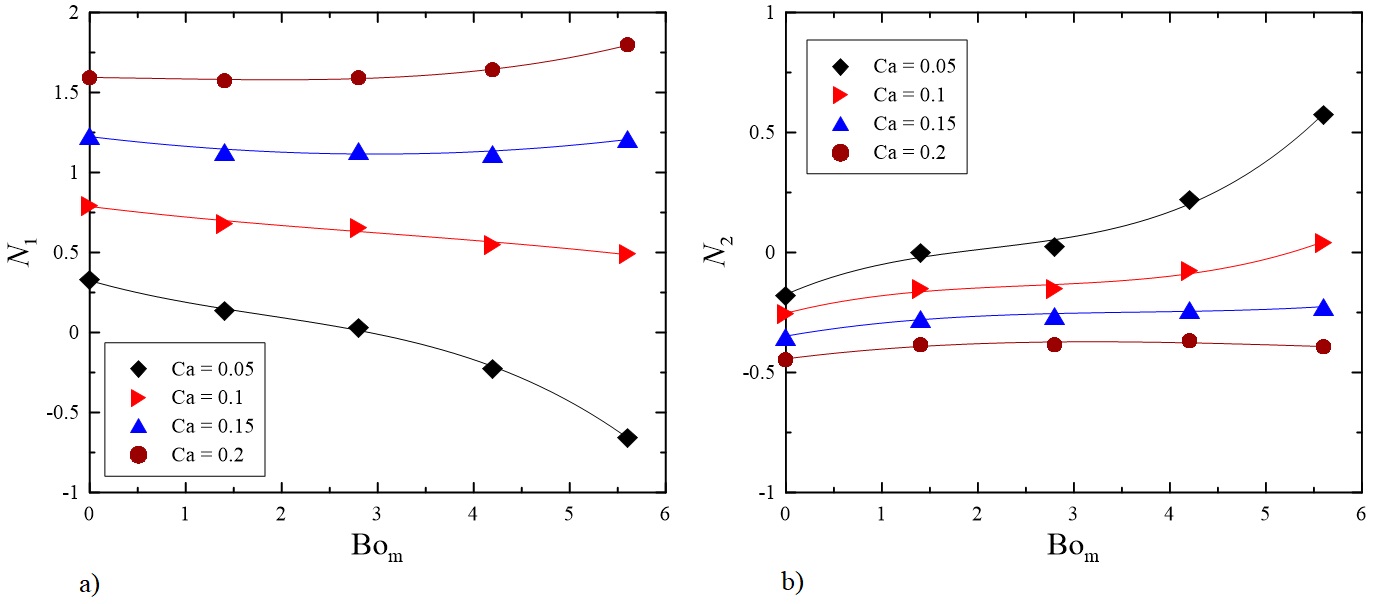}
\caption{\label{fig::normal_stresses_Bo}{First and second normal stress differences as a function of $\mathrm{Bo_m}$ for different $\mathrm{Ca}$. $\chi=0.5$. Data are scaled with the volume fraction $\phi$.}}
\end{figure}

We continue our discussion by showing the normal stress differences derived with our model. In this regard, in Fig. \ref{fig::normal_stresses} we report $N_1$, $N_2$ as a function of the capillary number {for different} $\mathrm{Bo_m}$. From these plots we immediately {realize that the general trend observed for both normal stress differences resemble those observed in the case $\mathrm{Bo_m}=0$. Nevertheless}, we note the presence of a reversal in the sign of both normal stresses {in the range of small $\mathrm{Ca}$ for the larger values of $\mathrm{Bo_m}$, i.e., for $\mathrm{Bo_m}=4.2,\,5.6$}. This is attributable to the fact that the magnetic torque, which counteracts the shearing of the imposed flow, forces the drop to be elongated and prominently oriented toward the vertical direction ($\theta'<0$) thereby introducing a stress anisotropy {enhanced toward the direction of the magnetic field}. Moreover, by a direct comparison with Fig. \ref{fig::validation}d, we observe that the extent of the normal stresses in the presence of magnetic field is in general {different} than that obtained in the absence of magnetic field. {The continuous lines added represent polynomial cubic fits.}

{The same set of results can also be displayed versus $\mathrm{Bo_m}$ for fixed capillary number, as shown in Fig. \ref{fig::normal_stresses_Bo}, from which we can draw some interesting considerations. Fig. \ref{fig::normal_stresses_Bo}a, in particular, shows the trends for $N_1$, and a comparison between the low- and high-$\mathrm{Ca}$ data, shows rather different behaviours. For the largest values of the capillary number, i.e., $\mathrm{Ca}=0.15$ and $\mathrm{Ca}=0.2$, we observe that $N_1$ increases with $\mathrm{Bo_m}$. On the contrary, for the remaining values of $\mathrm{Ca}$, the first normal stress difference decreases with $\mathrm{Bo_m}$. These opposite behaviours can be ascribed to the configuration assumed by the drop for different flow conditions stemming from the competition between magnetic and viscous forces. For increasing capillary numbers, in fact, we have seen that the orientation $\theta'$ increases monotonically. Therefore, for large values of $\mathrm{Ca}$, the anisotropy of the system is enhanced in the direction of the mean flow ($x$-direction), thereby promoting the increment of $N_1=\Sigma_{xx}-\Sigma_{yy}$ (note that for $\mathrm{Ca}=0.2$, $\theta'$ was found to be always positive). On the contrary, in the opposite case scenario small capillary numbers lead to a decrease of the orientation ($\theta'<0$), therefore the anisotropy of the system is enhanced in the direction of the magnetic field, thereby favouring the increment of the vertical normal stress $\Sigma_{yy}$ compared to the $\Sigma_{xx}$ component. This circumstance therefore leads to a progressive reduction of $N_1$ when $\mathrm{Bo_m}$ is increased. Regarding the results for $N_2$ shown in Fig. \ref{fig::normal_stresses_Bo}b, we observe that opposite considerations apply. When $\mathrm{Ca}$ is sufficiently small, $N_2$ increases with $\mathrm{Bo_m}$ due to the increase of $\Sigma_{yy}$, conversely, for increasing $\mathrm{Ca}$, $\Sigma_{yy}$ is progressively decreased and the increments of $N_2$ become less pronounced.}

\subsection{Rheological functions in the presence of magnetic field: comparison with existing theoretical models}
\label{sec::comparison_other_models}
{In the introductory section, it has been mentioned that Cunha et al. \cite{cunha2020} and Ishida and Matsunaga \cite{ishida2020}, have derived models for the same type of emulsion considered here using the approach detailed in Batchelor \cite{batchelor_1970}. However, these authors relied on the formulation introduced by Kennedy et al. \cite{KENNEDY1994251}, in which the quantity inside the surface integral appearing in Eq. (\ref{eq::particle_stress_general_reduced}) is rewritten as $x_j\Delta t_i$ upon the application of the divergence theorem, where $\Delta t_i$ is the interface traction jump. In this formulation, the stress was already reduced to the rightmost (volume) integral appearing in Eq. (\ref{eq::particle_stress_general_reduced}) upon the hypothesis of negligible inertia and uniform body force mentioned before (we recall that, contrarily, we found advisable retaining the body force term since the magnetic body force is not uniform throughout the flow domain). In spite of the fact that both Cunha et al. \cite{cunha2020} and Ishida and Matsunaga \cite{ishida2020} shared the same starting point, they followed different routes and came across different formulations for the magnetic extra stress tensor: in Cunha et al. \cite{cunha2020}, in fact, the magnetic extra-stress tensor is proportional to the square of the magnetic field intensity, $H^2$. On the other hand, Ishida and Matsunaga \cite{ishida2020} derived their model relying on the fact that in their numerical framework the interface is not sharp and therefore they approximated the surface integral of the tensor $x_j\Delta t_i$ with a volume integral evaluated over the finite thickness interfacial layer. As a result, their magnetic contribution to the particle stress tensor resembles the one derived in the present work (formulation in the framework of our numerical approach) with the difference of being transposed and having the opposite sign with respect to ours.} 

{A comparison with the stress calculation using the integral formulation detailed in Eqs. (\ref{eq::excess_stress_integral})-(\ref{eq::excess_shear_stress}) has shown the reliability of the formulations reported in Cunha et al. \cite{cunha2020} and Ishida and Matsunaga \cite{ishida2020} in providing accurate prediction of the total shear extra-stress, circumstance that has been encountered also with our model. Hence, we can argue that all models are capable to provide similar predictions of this quantity. Regarding the normal stress differences, Cunha et al. \cite{cunha2020} have reported positive increasing values for $N_1$ (their setup was two-dimensional, hence $N_2$ was not contemplated) for increasing capillary number. On the other hand, Ishida and Matsunaga \cite{ishida2020} considered three-dimensional configurations, but no reversal of the signs of $N_1$, $N_2$ were observed. Since their magnetic extra-stress tensor share a similar structure to the one determined in this work, being only transposed and changed in sign, the occurrence of a different behaviour in terms of normal stress differences can be expected.}
\begin{figure}[h!]
\centering
\includegraphics[width=0.95\textwidth]{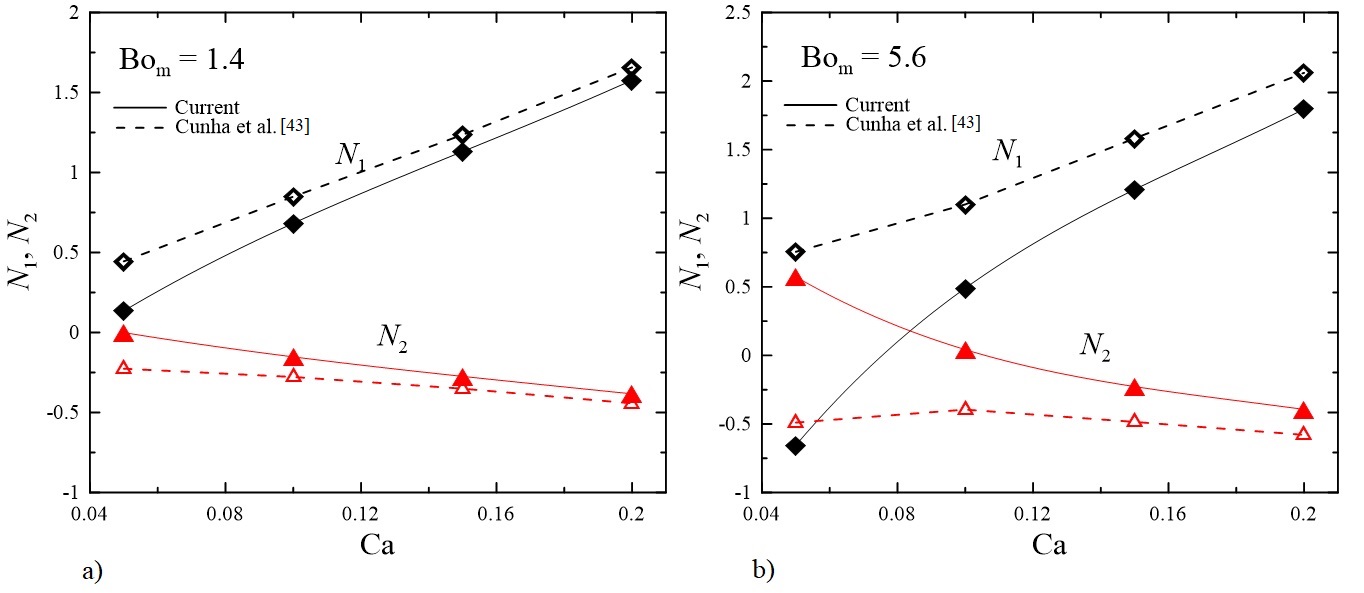}
\caption{\label{fig::normal_stresses_cunha} {Comparison in terms of first and second normal stress differences with the model of Cunha et al. \cite{cunha2020} as a function of $\mathrm{Ca}$ for $\mathrm{Bo_m=1.4}$, $\mathrm{Bo_m=5.6}$ and $\chi=0.5$. Data are scaled with the volume fraction $\phi$.}}
\end{figure}

{In Fig.\ref{fig::normal_stresses_cunha}, we show the normal stress differences vs capillary number calculated for our flow conditions adopting the model of Cunha et al. \cite{cunha2020} (dashed lines) for the cases $\mathrm{Bo_m}=1.4$ and $\mathrm{Bo_m}=5.6$ compared to our findings (continuous lines). It appears clear that both models predict similar trends, nevertheless, in line with the previous two-dimensional findings of Cunha et al. \cite{cunha2020}, also in these conditions reversal of normal stresses was not observed (cf. Fig. \ref{fig::normal_stresses_cunha}b). Completely analogous trends were found for the intermediate values of $\mathrm{Bo_m}$ which are not reported here for the sake of brevity. Regarding shear stresses, the model of Cunha et al. \cite{cunha2020} provided results essentially identical to those obtained with our model and shown in Fig. \ref{fig::sigma_excess}, with relative differences contained within 1\%.}

\section{\label{conclusions}conclusions}

The rheological properties of a dilute emulsion composed of ferrofluid droplets suspended in a non-magnetizable fluid have been investigated numerically considering uniform magnetic fields applied in the direction transverse to the imposed shear. Three-dimensional simulations have been carried out with a multiphase OpenFOAM code previously developed by Capobianchi et al. \cite{CAPOBIANCHI2018313} capable of dealing with interfacial flows in the presence of ferrofluid phases. A novel model for the bulk rheology of the emulsion based on the early work of Batchelor \cite{batchelor_1970}, has been derived assuming Newtonian behaviour for both phases, negligible inertia and linearly magnetizable fluids. 

The accuracy of the multiphase numerical framework in a three-dimensional setup has been initially tested in the absence of magnetic effects against the models of Choi and Schowalter \cite{choi1975} and Yu et al. \cite{yu_bousmina2002} for a isodense and isoviscous system. A general good agreement in terms of droplet morphology (deformation and orientation) and bulk rheology was found in a fairly broad range of capillary number. Subsequently, we imposed different moderate uniform magnetic fields while setting a magnetic susceptibility $\chi=0.5$, held constant throughout the whole study. In line with the previous two-dimensional calculations of Capobianhi et al. \cite{CAPOBIANCHI2018313} and Cunha et al. \cite{cunha2020}, {as well as with the three-dimensional computations of ishida and Matsunaga \cite{ishida2020}}, the droplet morphology was found to be significantly affected by the presence of magnetic stresses. In particular, it was found that magnetic effects contribute to enhance the drop deformation and orient it along the direction of the imposed magnetic field. Consequently, the rheological properties of the emulsion were found to be different to those observed for non-magnetizable fluids. In line with the non-magnetic case, the excess shear stress was found to be a monotonic decreasing function of $\mathrm{Ca}$ (for each value of $\mathrm{Bo_m}$), nevertheless the relative reduction of viscosity appeared to be less pronounced than the corresponding situation where the magnetic field was not considered. {Arguably, a larger magnetization could lead to the opposite scenario, i.e., to the appearance of a shear-thickening behavior.}

Calculation of the excess shear stresses obtained with the present model provided results that are in good agreement with the direct calculation of the stresses, thereby indicating the reliability of the present model. Calculations of the excess shear stress obtained with the model proposed by Cunha et al. \cite{cunha2020} also provided results in line with the present model in the whole range of parameters considered. Conversely, for constant values of $\mathrm{Ca}$, the excess shear stresses were found to be a monotonic increasing function of $\mathrm{Bo_m}$ (magneto-thickening behaviour), and the available data suggested a {cubic} dependence with this latter parameter. On the basis of this observation, we proposed a simple constitutive equation for the emulsion describing its viscosity as a function of the applied magnetic field, i.e., as a function of $\mathrm{Bo_m}$ while keeping constant the imposed shear.

In terms of normal stresses, our model predicted a reversal of the sign of both first and second normal stress differences with respect to those obtained for the non-magnetizable case {for those conditions in which the imposed magnetic force prevails over the viscous force}. We concluded that such behaviour can be ascribed to the strong anisotropy introduced by the magnetic stresses which contribute to deform and orient the drop toward the direction of the magnetic field. 

\section*{Declaration of interests}
The authors report non declaration of interests.

\section*{Supplementary information}
See the supplementary {information} for the complete validation study of the present numerical framework.

\section*{Acknowledgments}
FTP wishes to thank financial support provided by Centro de Estudos de Fenómenos de Transporte through projects UIDB/00532/2020 and UIDP/00532/2020. Oliveira acknowledges funding from the Glasgow Research Partnership in Engineering (GRPe).

\appendix

\section{The C-S and the GBP-YB Models}

Here we report the C-S  model of \cite{choi1975} and the GBP-YB model of \cite{yu_bousmina2002} adopted in the validation section for the convenience of the reader.
\subsection{The C-S Model}
Choi and Schowalter \cite{choi1975} developed a rheological model for emulsion in steady shear Stokes flow based on the small deformation perturbation analysis. As the volume fraction $\phi\to0$, the rheological functions vary linearly with volume fraction $\phi$. Consider the viscosity ratio $\lambda = 1$, the interfacial rheological functions are reduced to:
\begin{equation}
    \Sigma_{xy}^{\mathrm{C-S}}=\frac{\eta_e^{\mathrm{C-S}}}{\eta}=\frac{7}{4}\phi,
\end{equation}
\begin{equation}
    N_1^{\mathrm{C-S}}=\frac{245}{32}\frac{\mathrm{Ca}}{\left(1+Z^2 \right)}\phi,
\end{equation}
\begin{equation}
    N_2^{\mathrm{C-S}}=-\frac{35}{16}\frac{\mathrm{Ca}}{\left(1+Z^2 \right)}\phi,
\end{equation}
where
\begin{equation}
    Z=\frac{35}{16}\mathrm{Ca}.
\end{equation}
\subsection{The GBP-YB Model}
Based on Grmela et al. \cite{garmelaetal2001} morphological tensor model, Yu et al. \cite{yu_bousmina2002} calculated the interfacial rheological functions for emulsion in shear Stokes flow. Shear-rate dependence of viscosity is taken into account. The expressions for these functions are:
\begin{equation}
    \Sigma_{xy}^{\mathrm{GBP-YB} }=\frac{\eta_e^{\mathrm{GBP-YB}}}{\eta}=\frac{128}{35}\frac{\phi}{S},
\end{equation}
\begin{equation}
    N_1^{\mathrm{GBP-YB}}=16\frac{\phi}{S}\mathrm{Ca},
\end{equation}
\begin{equation}
    N_2^{\mathrm{GBP-YB}}=-\frac{1}{2}N_1^{\mathrm{GBP-YB}},
\end{equation}
where
\begin{equation}
    S=\left(10-7\phi \right)\left(\mathrm{Ca}^2+\frac{256}{1225} \right).
\end{equation}

\section{Ferrofluid code validation}
\label{sect::appendixB}
\begin{figure}
\centering
\includegraphics[width=0.95\textwidth]{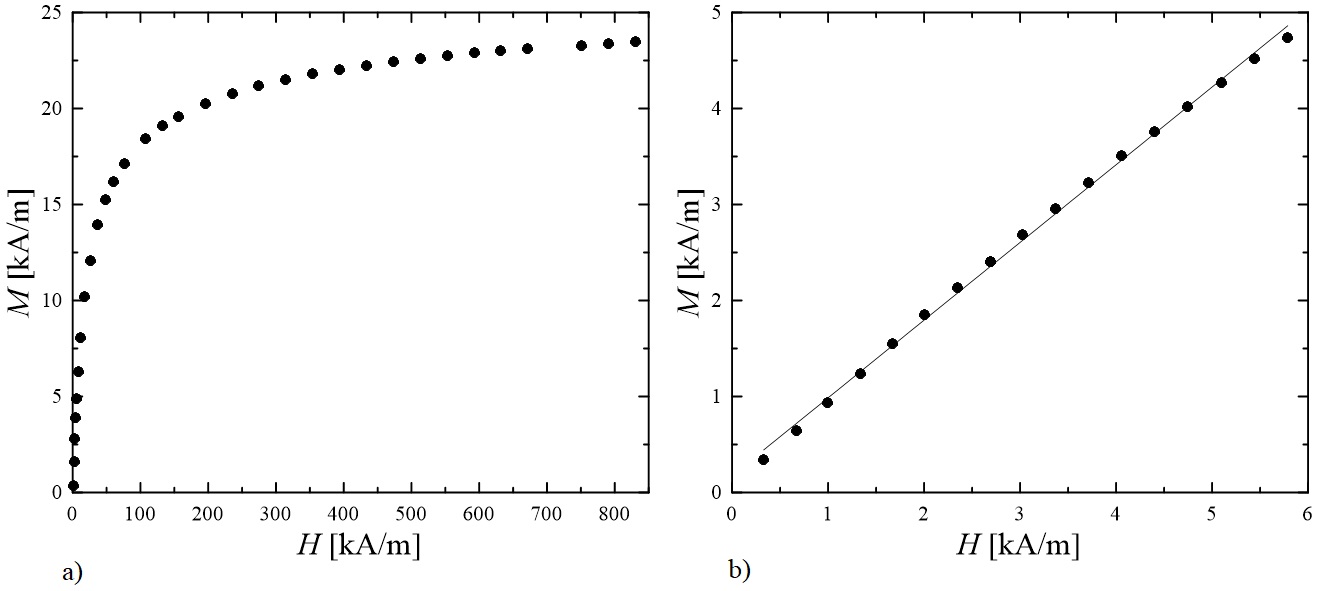}
\caption{\label{fig::langevin}{High (a) and low (b) field magnetization curves for the  7 vol. \% magnetite ($\mathrm{Fe}_3\mathrm{O}_4$) particles with a mean diameter of $7.2\,\si{nm}$ dispersed in glycerol ($\mu_{glyc}\simeq\mu_0$) determined by \cite{afkhami2010}.}}
\end{figure}

In this Appendix, we report the {the magnetization (high- and low-field intensity) curves reported in \cite{afkhami2010} which are also relevant for the present study, as we have referred to the same type of ferrofluid for our numerical simulations,} and the validation of the code developed by \cite{CAPOBIANCHI2018313} {against the experiments of \cite{afkhami2010}}. In this regard, {in Fig. \ref{fig::langevin}a,b we report the results of the measurements of \cite{afkhami2010} for the magnetization for both high (Fig. \ref{fig::langevin}a) and low (Fig. \ref{fig::langevin}b) magnetic field intensity. In our numerical simulations, the applied magnetic field was always constant and set to a value $H=750\,\si{A/m}$, i.e., within the limit of the small magnetization curve. Nevertheless, we should observe that in an actual emulsion the dimension of the drops are expected to be several order of magnitude smaller than that considered here. Thus, it is necessary to check whether for an emulsion with droplets having reasonably small size, the values of the relative magnetic Bond number are reasonably large for the intensity of the magnetic field that are in the limit of low fields intensity. Thus, considering the largest magnetic field reported in Fig. \ref{fig::langevin}b, i.e., $H\approx 6\,\si{kA/m}$, and the interfacial tension the value of $\sigma\approx 10\,\si{mN/m}$ (cf. Refs. \cite{,afkhami2010,rowghanian_2016}), we infer that the droplet dimension should be on the range $O\left(10^{-4}\right)\,\si{m}$ to $O\left(10^{-3}\right)\,\si{m}$ to obtain the order of magnitude of the magnetic Bond numbers considered in this work. With a fluid pair having a smaller interfacial tension, see for instance \cite{flament1996,Zakinyan2011}, smaller droplet sizes would lead to similar values of $\mathrm{Bo_m}$. For instance, Zakinyan and Dikansky \cite{Zakinyan2011} reported a value of the interfacial tension, $\sigma=10^{-3}\,\si{mN/m}$, for their system composed by drops of a kerosene-based ferrofluid dispersed in a FH51 aviation oil. They were able to produce significant displacement of micron-sized drops with the application of relatively low magnetic fields (order of few \si{\kilo \ampere/\metre} or smaller).} 

\begin{figure}
\centering
\includegraphics[width=0.5\textwidth]{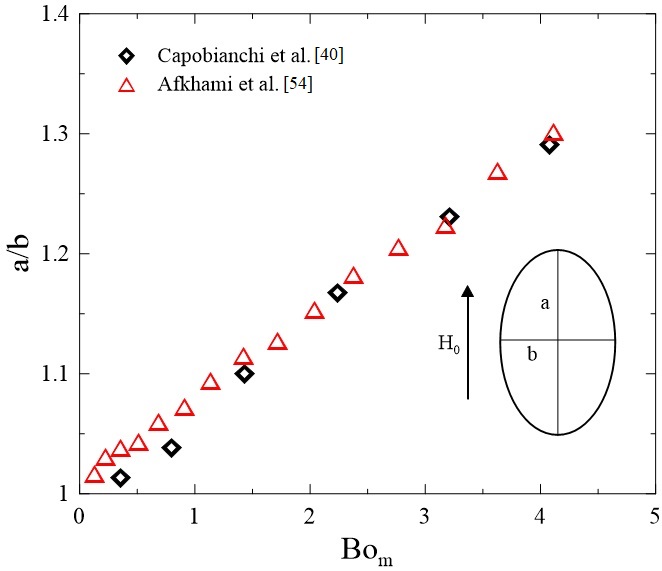}
\caption{\label{fig::validation_ff}Deformation of a ferrofluid droplet immersed on a non-magnetizable fluid as a function of the magnetic Bond number. Comparison between the numerical results of \cite{CAPOBIANCHI2018313} and the experiments of \cite{afkhami2010}.}
\end{figure}

Fig. \ref{fig::validation_ff} shows the deformation of a ferrofluid drop surrounded by a non-magnetizable fluid measured as the ratio between the major and minor axes (refer to the inset). The red symbols are representative of the experiments of \cite{afkhami2010}, while the black ones are the simulation carried out by \cite{CAPOBIANCHI2018313}. The value of the magnetic susceptibility was $\chi=0.8903$, as reported by the measurements of \cite{afkhami2010}. Apart from a small discrepancy at the low-$\mathrm{Bo_m}$ regime ($\mathrm{Bo_m}<1$), which can be attributed to the aforementioned problem related to the interface-capturing numerical methodology adopted here, the two sets of results are in fairly good agreement.

\newpage
\printbibliography

\end{document}